\documentclass[aps,pra,preprint,superscriptaddress, nofootinbib,unsortedaddress]{revtex4-1}
\usepackage{amsmath}
\usepackage{graphicx}
\usepackage{amssymb}
\usepackage{amsthm}
\usepackage{amscd}
\usepackage{dsfont}
\usepackage{verbatim}
\usepackage{color}
\numberwithin{equation}{section}
 \DeclareMathOperator{\Tr}{\rm{Tr}}

\begin{document}
\title{Algebraic Approach to Entanglement and Entropy}
\author{A. P. Balachandran}
\email[]{balachandran38@gmail.com}
\affiliation{Institute of Mathematical Sciences, CIT Campus, Taramani,
Chennai
600113, India}
\affiliation{Physics Department, Syracuse University, Syracuse, NY,
13244-1130,
USA}

\author{T. R. Govindarajan}
\email[]{trg@imsc.res.in}
\affiliation{Institute of Mathematical Sciences, CIT Campus, Taramani,
Chennai
600113, India}

\affiliation{Chennai Mathematical Institute, H1, SIPCOT IT Park,
Kelambakkam,
Siruseri 603103, India}

\author{Amilcar R. de Queiroz}
\email[]{amilcarq@unb.br}
\affiliation{Instituto de Fisica, Universidade de Brasilia, Caixa Postal
04455,
70919-970, Brasilia, DF, Brazil}
\altaffiliation{Institute of Mathematical Sciences, CIT Campus, Taramani,
Chennai 600113, India}

\author{A. F. Reyes-Lega}
\email[]{anreyes@uniandes.edu.co}
\homepage[]{http://fisicateorica.uniandes.edu.co/anreyes/}
\affiliation{Departamento de F\'isica, Universidad de los Andes, A.A.
4976,
Bogot\'a D.C., Colombia}
\altaffiliation{Institute of Mathematical Sciences, CIT Campus, Taramani,
Chennai 600113, India}
 
\date{\today}

\begin{abstract}
We present a general approach to quantum  entanglement and entropy that is based on algebras of observables and states thereon. In contrast to more standard treatments, Hilbert space is an emergent concept, appearing as a representation space of the observable algebra, once a state is chosen. In this approach, which is based on the Gel'fand-Naimark-Segal construction, the study of subsystems becomes particularly clear. We explicitly show how the problems associated with partial trace for the study of entanglement of identical particles are readily overcome. In particular,  a suitable entanglement measure is proposed, that can be applied to systems of particles obeying Fermi, Bose, para- and even braid group statistics. The generality of the method is also  illustrated by the study of time evolution of subsystems emerging from restriction to subalgebras. Also, problems related to anomalies and quantum epistemology are discussed.
\end{abstract}
\pacs{03.67.Mn, 03.65.Ud, 89.70.Cf, 02.30.Tb} \keywords{} \maketitle
\section{Introduction}
\label{sec:Introduction}
\subsection{On States and Algebras}

In a physical theory, the theoretical description of observations contains two ingredients. The first is the state of the
system being observed, it contains the data specifying the  system. The second is the specific observable of the system
being measured. The output of observations involves a suitable pairing of the state to the observable which yields a
number.

In both classical and quantum physics, the state $\omega$  provides a probability distribution and observables form an
(associative) algebra.

In classical physics, the state $\omega_c$ gives a probability distribution in phase space whereas the algebra $\mathcal
A_c$ of observables consists of real-valued functions on phase space\footnote{More generally one talks of observables as ``hermitean'' functions on a complexified phase space.}. The product of two functions $\alpha,\beta$ on
phase space is point-wise multiplication: for a point $x=(q,p)$ in phase space, $(\alpha\beta)(x)= \alpha (x) \beta(x)$. The algebra $\mathcal A_c$ is commutative.

The pairing of $\omega_c$ to $\alpha$ produces the mean value of $\alpha$, as follows. The mean value of $\alpha$ for the state $\omega_c$ is
\begin{eqnarray}
\omega_c(\alpha) & = & \int d\mu(x)~\rho_c(x)\alpha(x), \\
d\mu(p) & = & \mbox{Liouville measure on phase space},
\end{eqnarray}
and $\rho_c$ is the probability density on phase space associated with $\omega_c$.

We note that being a probability density, $\rho_c$ is normalised as
\begin{equation}
\omega_c(\mathds 1) \equiv \int d\mu(x)\rho_c(x)=1,
\end{equation}
where $\mathds 1$ is the constant function with value $1$:
\begin{equation}
\mathds 1 (x)=1.
\end{equation}

In quantum theory we still have a state $\omega$ and an algebra $\mathcal A$. The state $\omega$ on an observable $\alpha\in\mathcal{A}$ is generally representable in terms of a density matrix $\rho_\omega$ and an operator $\pi_\omega(\alpha)$ representing $\alpha$ on a Hilbert space. The mean value of the observable $\alpha$ is then
\begin{equation}
\omega(\alpha)= \Tr \left(\rho_\omega \alpha)\equiv \Tr(\rho_\omega\pi_\omega(\alpha) \right).
\end{equation}
From now on, we represent $\pi_\omega(\alpha)$ by $\alpha$ itself, if there is no ambiguity. The state and its density matrix are normalised just as in the classical case:
\begin{equation}
\omega(\mathds 1)=\Tr \rho_\omega = 1.
\end{equation}

The basic mathematical  difference between classical and quantum physics lies in this: \emph{classical physics is a  probability theory
on a commutative algebra $\mathcal A_c$, quantum physics is a probability theory on a non-commutative algebra $\mathcal A$}.

A state $\omega$ need not be presented using  a density matrix to extract numbers from theory. It is enough that $\omega$ is a linear map from $\mathcal A$ to $\mathds C$ with the properties
\begin{equation}
\label{state-1}
\omega(\mathds 1)= 1, \quad \omega(\alpha^*)=\overline{\omega(\alpha)}, \quad \omega(\alpha^*\alpha) \geq 0 \qquad \mbox{for all}\;\;\alpha\;\;\mbox{in}\;\;\mathcal A,
\end{equation}
where ``$*$'' is a hermitean conjugation (``anti-linear involution'') $\mathcal A$ should have. A state is best understood in this manner.
State vectors and Hilbert spaces play no role at this point. Where, then, do they come from?

\subsection{On the GNS Construction}

State vectors and Hilbert spaces are best thought of as emergent concepts in quantum physics. The primary concepts
are states $\omega$ and the algebra $\mathcal A$ of observables.

In the 1940's, Gel'fand, Naimark and Segal described  the reconstruction of the Hilbert space $\mathcal H_\omega$ from the data $(\mathcal A, \omega)$. The algebra $\mathcal A$ acts by a representation $\pi_\omega$ on $\mathcal H_\omega$. This reconstruction, known as the GNS construction, has played a foundational role in the theory of operator algebras. It has also been an important tool for studies in quantum field theory \cite{haag1996local}.

We suggest in this paper that the GNS construction is the proper framework for the study of entanglement as well. A brief account of our ideas has appeared before \cite{balachandran2012a}.

In section \ref{sec:2}, we describe the GNS reconstruction of $(\mathcal H_\omega, \pi_\omega)$ from $(\mathcal A, \omega)$. We do not aspire to rigour but to concepts and computations. Whatever we say is correct in finite dimensions. But our presentation omits the fine points of topology for infinite dimensional $\mathcal A$.

The GNS construction presents the state $\omega$ as a density matrix $\rho$. We can expand $\rho$ in terms of orthogonal rank 1 density matrices $\rho_i$:
    \begin{equation}
     \rho_i\rho_j=\delta_{ij}\rho_i, \quad \Tr \rho_i = 1,
    \end{equation}
and accordingly write
\begin{equation}
\label{eq:1.9}
\rho=\sum_i \lambda_i\rho_i,\;\;\;\;\omega=\sum_i\lambda_i\omega_i,\;\;\;\lambda_i>0,\;\;\;\sum_i\lambda_i=1,
\end{equation}
where $\omega_i$  is the state associated with the density matrix $\rho_i$. The von Neumann entropy for $\omega$ is then
\begin{equation}
\label{eq:1.10}
S(\omega)=-\Tr~\rho\log\rho=-\sum_i\lambda_i\log\lambda_i.
\end{equation}
We can thus associate an entropy to a pair $(\mathcal A, \omega)$ of a state and an algebra of observables. This result is important for us.

\subsection{What is entanglement}

For a system of \emph{non-identical} constituents $A_i$ with Hilbert spaces $\mathcal H_{i}$, `entanglement' can be understood in terms of `partial trace' as follows.

It is enough to consider a bipartite system and the Hilbert space
\begin{equation}
\mathcal H=\mathcal H_{1}\otimes \mathcal H_{2}.
\end{equation}
Then a density matrix
\begin{equation}
 \rho_{12}=|\psi\rangle\langle\psi|
\end{equation}
with a normalised  vector state $|\psi\rangle\in \mathcal H$ is entangled when the density matrix
\begin{equation}
\rho_{i}=\Tr_{j} \rho_{12} \;\;\;\;(j,i=1,2, \quad i\neq j)
\end{equation}
obtained by partial tracing has non-zero von Neumann entropy $S(\rho_{i})$:
\begin{equation}
S(\rho_{i})= -\Tr \rho_{i} \log \rho_{i}\nonumber \neq 0.
\end{equation}
There is no entanglement if $S(\rho_{i})=0$.

In section \ref{sec:3}, we recall that partial tracing has poor physical meaning for systems of \emph{identical} particles. That is evident from the impressive volume of inconclusive literature on fermions and bosons, and the apparent absence of any treatment of para- and braid- statistics~\cite{Li2001,Paskauskas2001,Schliemann2001,Zanardi2002,Fang2003,Ghirardi2004,L'evay2005,Banuls2009, Plastino2009, Zander2012}.

It is thus desirable to replace ``partial tracing'' by better concepts to study entanglement.

In this section we reformulate the notion of partial tracing in terms of the restriction $\omega_0:= \omega|_{\mathcal A_0}$ of a state $\omega$ on an algebra $\mathcal A$ to a subalgebra $\mathcal A_0$. The example we choose is very simple: the bipartite system of non-identical particles. This restatement is possible whenever partial tracing can be understood in terms of a restriction. When such interpretation is not possible, partial tracing loses meaning. But the restriction of $\omega$ to subalgebras is always sensible.

The notion of ``local operation'' (as used in quantum information theory) is implemented here through the choice of an appropriate subalgebra $\mathcal A_0 \subseteq \mathcal A$. Importance of focusing on subsystems using subalgebras has also been emphasized in  \cite{Barnum2004,Benatti2012, Harshman2011,Derkacz2012}.  In our approach,  the restriction of a state $\omega$ on $\mathcal A$ to the subalgebra $\mathcal A_0$ will give rise to a ``reduced'', or restricted state $\omega_0$, whose entropy compared to that of $\omega$ provides a measure of the entanglement of $\mathcal A_0$ with $\mathcal A$ in the state $\omega$.

For the purposes of this paper, this is what is meant by entanglement.

If $\omega$ is pure, that is, $S(\omega)=0$, the entropy of entanglement is obtained from the expansion
\begin{equation}
\omega_0= \sum_i \lambda_i \omega_{0,i}\;\;\;\lambda_i>0\;\;\;\sum_i\lambda_i=1
\end{equation}
of $\omega_0$ in terms of pure states and using the formula
   (\ref{eq:1.10}). There are orderly ways to calculate the expansion (\ref{eq:1.9})  and that will be taken up
   in the following sections.

If $\omega$ itself is not pure, we can compare its entropy  with that of $\omega_0$.

\subsection{Examples}

   We illustrate  the GNS approach to entropy by several examples in section \ref{sec:4}. The first is simple and involves just the matrix algebra $\mathcal A= M_2(\mathds C)$.

This example is followed up using those from identical fermions and bosons. We clarify what is meant by observing single particle operators of a boson or a fermion in a two-particle system. For this purpose, we need at the simplest level that the single-particle algebra $\mathcal A$ is Hopf with a coproduct $\Delta$
compatible with the permutation group which here defines statistics\footnote{It is perhaps enough if $\mathcal{A}$ is quasi-Hopf or just a coalgebra.} \cite{Balachandran2010}.
This $\Delta$ allows us to identify the correct `single particle' subalgebra $\Delta(\mathcal A)$ of the full two-particle algebra and calculate
$\omega|_{\Delta(\mathcal A)}$ and its entropy for any two-body state $\omega$. It does not agree with what one finds from partial tracing.

We can also illustrate all this for para-bosons and para-fermions. For these cases, $\Delta$ is not changed. So we instead consider the braid group $B_N$ to be the $N$-particle statistics group. In that case $\Delta$ is changed to $\Delta_b$. We calculate the restriction of a chosen three-particle state to single-particle observables, obtained from $\Delta_b$ in a systematic manner and calculate its entropy. It illustrates the effectiveness of the GNS approach when combined with concepts from Hopf algebras, in quantum information theory.

If the subalgebra $\mathcal A_0$ comes from a $k$-particle  subsystem  in an $N$-particle system, it is necessary that $\mathcal A_0$ is at least a coalgebra with a coproduct $\Delta$  to extend its meaning beyond $N$-particles. \emph{This is a new point of our work}.

Second quantization also requires the notion of coproduct \cite{Balachandran2010}.

   \subsection{Time evolution}

   In the GNS-approach, given the time evolution $\omega(t)$ of a state $\omega\equiv \omega(0)$ on $\mathcal A$, it induces a time evolution on the restriction $\omega_0=\omega|_{\mathcal A_0}$ of $\omega$ to $\mathcal A_0$. It is
   \begin{equation}
   \omega_0(t)=\omega (t)|_{\mathcal A_0}.
   \end{equation}
   If $\omega(0)$ is represented by a density matrix $\rho(0)$ with unitary time evolution according to
   \begin{eqnarray}
   \rho(t) &=& U(t)^{-1}\rho(0) U(t),\nonumber\\
   U(t)&=&e^{-itH}, \;\;\; H=\mbox{Hamiltonian},
   \end{eqnarray}
   there is no assurance that $\rho_0(t)=\rho(t)|_{\mathcal A_0}$ also undergoes a unitary evolution.

   The generic situation will rather be that of an evolution by positive maps. This follows from Stinespring's  theorem \cite{Stinespring1955}.

   We illustrate the evolution $\rho_0(t)\rightarrow \rho_0(t+\tau)$ by completely positive maps using our algebraic approach, in section \ref{sec:5}. Here we only remark that this type of evolution arises very naturally from the restriction to a subalgebra. This is the basic mechanism behind quantum decoherence.

   \subsection{Anomalies}

It has been suggested \cite{Balachandran2012,Balachandran2012b} that unwanted anomalies, such as the $T$-violation associated with the QCD $\theta$-angle, can be eliminated by using mixed states.
The appearance of such mixed states for restoring anomalies can also be explained in terms of restrictions of states to subalgebras, as we show in section \ref{sec:6}. This argument provides further strong evidence for the power of the proposition for the use of mixed states in that context.

The GNS theory is thus useful in many different physical contexts.

\subsection{On state restrictions and quantum epistemology}

We suggest in section \ref{sec:7} of this paper that the restriction $\omega_0=\omega|_{\mathcal A_0}$ of a state $\omega$ on $\mathcal A$ can also be understood in terms of
``collapse of wave packets'' and an interpretative superstructure of quantum theory. Thus suppose that we have an algebra $\mathcal A$ of observables acting on a Hilbert space $\mathcal H$ and we observe a projector $p\in \mathcal A$. Let $\mathcal A_0\subset \mathcal A$ be the maximal subalgebra which commutes with $p$ (and hence includes $p$).
 Then we can interpret $\omega_0$ as the restriction of $\omega$ to $\mathcal A_0$. We explain how this is so in that section.

\section{The GNS Construction}
\label{sec:2}

For completeness, we begin with a few definitions. They are all effortlessly fulfilled in all our examples where we consider only finite-dimensional matrix algebras.

A $*$-algebra $\mathcal A$ is an associative algebra over $\mathds C$ with an anti-linear involution (``hermitean conjugation''):
    \begin{align}
     *:\mathcal{A}&\rightarrow \mathcal{A} \nonumber \\
     *^2 &= \textrm{id}.
    \end{align}

A $C^*$-norm $\|\cdot\|$ on such an algebra $\mathcal A$ is a norm fulfilling the property
\begin{equation}
\|\alpha^* \alpha\|=\|\alpha\|^2, \;\;\;\forall \alpha\in \mathcal A.
\end{equation}
If it exists, it is unique.

Algebras $M_d(\mathds C)$ of $d\times d$ matrices can all be regarded as $C^*$-algebras, with the norm fixed by
\begin{equation}
\|\alpha\|^2= \;\;\mbox{largest eigenvalue of}\;\; \alpha^*\alpha.
\end{equation}
The algebra $\mathcal A$ of observables in quantum theory is a $C^*$-algebra. Indeed, the algebra of all bounded operators $\mathcal B(\mathcal H)$ on a separable Hilbert space admits a norm to turn it into a $C^*$-algebra.

We will assume that $\mathcal A$ is unital with unity $\mathds 1_{\mathcal A}$. That is needed to discuss completeness relations, for example.

The data we are given is thus $(\mathcal A, \omega)$. We can construct $(\mathcal H_\omega,\pi_\omega(\mathcal A))$
from this data as follows.

For each element $\alpha \in \mathcal A$, we associate a vector $|\alpha\rangle$ in a complex vector space $\hat{\mathcal A}$ with the property
\begin{align}
|\lambda \alpha +\mu \beta\rangle & =  \lambda |\alpha\rangle +\mu |\beta\rangle, \\
\lambda,\mu\in \mathds C &;  \alpha,\beta \in \mathcal A \nonumber.
\end{align}
Next, an inner product is introduced in $\hat{\mathcal A}$ using $\omega$:
\begin{equation}
\label{eq:2.3}
\langle \beta | \alpha\rangle = \omega (\beta^* \alpha).
\end{equation}
It fulfills
\begin{equation}
\langle \mathds 1_{\mathcal A}|\alpha\rangle= \omega (\alpha)=\overline{\langle \alpha|\mathds 1_{\mathcal A}\rangle}= \overline{\omega (\alpha^*)}
\end{equation}
and Schwarz inequality as well in view of (\ref{state-1}) ($\omega(\alpha^*\alpha)\geq 0$):
\begin{equation}
|\langle \beta|\alpha\rangle|^2\leq \langle \beta|\beta\rangle\langle \alpha|\alpha\rangle.
\end{equation}
But it may not be a scalar product from which we can build a Hilbert space, as there may be $0\neq\alpha\in \mathcal A$ giving vectors $|\alpha\rangle$ of zero norm:
\begin{equation}
\langle\alpha|\alpha\rangle=0.
\end{equation}
Let $\mathcal N_\omega$ denote the subspace of $\mathcal A$ whose image $\hat{\mathcal N_\omega}\subset \hat{ \mathcal A}$ are vectors of zero norm:
\begin{equation}
\mathcal N_\omega=\lbrace \alpha\in  \mathcal A\;|\; \langle\alpha|\alpha\rangle=0\rbrace.
\end{equation}
Observe that, from Schwarz inequality,
\begin{equation}
\label{eq:2.7}
\langle a|\alpha\rangle=0 \;,\;\;\forall a\in \mathcal A,\alpha\in \mathcal N_\omega.
\end{equation}
Hence, $\mathcal N_\omega$ is a left-ideal. That is,
\begin{equation}
\label{eq:property-(ii)}
a\mathcal N_\omega \subseteq \mathcal N_\omega\;\;\;\forall a\in \mathcal A.
\end{equation}
This follows from (\ref{eq:2.7}): if $a\in \mathcal A$ and $\alpha\in \mathcal N_\omega$,
$\langle a\alpha|a \alpha\rangle=\langle a^* a \alpha|\alpha\rangle=0$ by (\ref{eq:2.7}). The subspace $\mathcal N_\omega$ is called the Gel'fand ideal.

We next consider the vector space
\begin{equation}
\hat {\mathcal A}/\hat{\mathcal N_\omega} = \lbrace |[\,a]\rangle:=|a+\mathcal N_\omega\rangle, \;\;a\in\mathcal A \rbrace.
\end{equation}
The label $[a]$ of a vector denotes an equivalence class $a+\mathcal N_\omega$, a set, in $\mathcal A$.

Now,
\begin{itemize}
\item[a)] $\hat{\mathcal A}/\hat{\mathcal N_\omega}$ has a well-defined  scalar product $\langle\cdot|\cdot\rangle$ given by
    \begin{equation}
    \label{eq:inner-prod}
    \langle [ a ] | [ b ] \rangle = \omega(a^* b).
    \end{equation}
In particular  the vector $|\mathcal N_\omega\rangle$, playing the role of zero, is the only vector that has zero norm. Note in this connection that  since $\omega (a^* (b+\alpha)) = \omega( a^*  b )$ for all $\alpha \in \mathcal N_\omega$, the right hand side of
(\ref{eq:inner-prod}) does not depend on the choices of representatives  $a,b$ from $[a]$ and $[b]$.

We denote the Hilbert space obtained by the closure of $\hat{\mathcal A}/\hat{\mathcal N_\omega}$ under (\ref{eq:inner-prod}) as $\mathcal H_\omega$.

\item[b)] Because of (\ref{eq:property-(ii)}), $\mathcal H_\omega$ carries a representation $\pi_\omega$ of $\mathcal A$:
\begin{equation}
\pi_\omega(a) |[b]\rangle:= |[ab]\rangle.
\end{equation}
\end{itemize}
We have now obtained $(\mathcal H_\omega, \pi_\omega(\mathcal A))$ from $(\mathcal A, \omega)$ thereby completing the GNS construction.

\subsection{Properties of the GNS Representation}

Consider the vector $|[\mathds 1_{\mathcal A}]\rangle$. Then, if $\pi_\omega\left(\mathcal{A}\right)$ denotes the set $\{\pi_\omega(a):~\forall a\in\mathcal{A} \}$,
\begin{equation}
\pi_\omega(\mathcal A)|[ \mathds 1_{\mathcal A} ]\rangle= \lbrace|[a]\rangle\;|\; a\in \mathcal A \rbrace,
\end{equation}
so that from $|[ \mathds 1_{\mathcal A} ]\rangle$ we can generate all vectors of $\mathcal H_\omega$ by acting with $\pi_\omega(\mathcal A)$ and closure. Such a vector $|[ \mathds 1_{\mathcal A} ]\rangle$ is said to be \emph{cyclic}. The representation $\pi_\omega$ is a \emph{cyclic} representation.

The state $\omega$ can now be represented as a density matrix $\rho_\omega$,
\begin{equation}
\rho_\omega = |[ \mathds 1_{\mathcal A} ]\rangle\langle[ \mathds 1_{\mathcal A} ]|.
\end{equation}
That is because
\begin{eqnarray}
\omega(a^* b ) & = & \langle[a]|[b]\rangle = \langle [ \mathds 1_{\mathcal A} ]|\pi_\omega(a^*) \pi_\omega(b)|[ \mathds 1_{\mathcal A} ]\rangle\nonumber\\
&=& \Tr \left(\rho_\omega \pi_\omega(a^*)\pi_\omega(b)\right).\end{eqnarray}

\subsection{On irreducibility and entropy}

The representation $\pi_\omega$ may not be irreducible. In finite dimensions at least, which are our concern here, it can be reduced to a direct sum of irreducible representations (IRR's) $\pi_\omega^{(\alpha)}$:
\begin{equation}
    \label{reducibility-1}
\pi_\omega = \oplus_\alpha \pi_\omega^{(\alpha)}.
\end{equation}
That is because $\mathcal A$ is a $*$-algebra. The proof is similar to the one for finite groups $G$,  their group algebras also being $*$-algebras.

The proof of (\ref{reducibility-1}) goes as follows. If $\mathcal H^{(1)}_\omega\subset \mathcal H_\omega$ is a non-trivial invariant subspace under $\pi_\omega(\mathcal A)$, then so is its orthogonal complement $\mathcal H^{(1)\perp}_\omega$. For if
\begin{equation}
|[\alpha]\rangle \in \mathcal{H}^{(1)}_\omega,\quad |[\beta]\rangle \in\mathcal{H}^{(1)\perp}_\omega,
\end{equation}
so that  $\langle[\alpha]|[\beta]\rangle= 0$, then
\begin{equation}
\langle[\alpha]|[a \beta]\rangle=0,\;\;\;\forall a\in \mathcal A,
\end{equation}
for the left-hand side is $\langle[a^*\alpha]|[\beta]\rangle$. But $a^*\in \mathcal A$ ($\mathcal A$ being $*$) and hence
$|[a^* \alpha]\rangle\in \mathcal H^{(1)}_\omega$. The statement follows.

Now, $\mathcal H^{(1)}_\omega$ (and similarly $\mathcal H^{(1)\perp}_\omega$) is either irreducible or has a non-trivial invariant subspace. If the latter is the case, we repeat the process ending up with
\begin{equation}
\mathcal H_\omega = \oplus_\alpha \mathcal H^{(\alpha)}_\omega,
\end{equation}
where the sum is the orthogonal direct sum and $\mathcal H^{(\alpha)}_\omega$ carries $\pi_\omega^{(\alpha)}(\mathcal A)$.

We now show how to correspondingly decompose $\rho_\omega$ into a convex sum of orthogonal rank 1 density matrices:
\begin{align}
\rho_\omega &=  \sum_\alpha \lambda_\alpha \rho_\omega^{(\alpha)},\;\;\;\lambda_\alpha>0,\;\;\sum_\alpha\lambda_\alpha=1,\nonumber\\
\rho_\omega^{(\alpha)} \rho_\omega^{(\beta)} &= \delta_{\alpha\beta} \rho_\omega^{(\alpha)}.
\end{align}
For this purpose, we write
\begin{equation}
|[\mathds 1_{\mathcal A}]\rangle = \sum_\alpha|[\mathds 1^{(\alpha)}_{\mathcal A}]\rangle,\;\;\;
|[\mathds 1^{(\alpha)}_{\mathcal A}]\rangle\in \mathcal H_\omega^{(\alpha)},
\end{equation}
that is, we decompose the left hand side into its components in $\mathcal H_\omega^{(\alpha)}$. Now, since for $\alpha\neq\beta$
\begin{equation}
\langle[\mathds 1^{(\alpha)}_{\mathcal A}]| \pi_\omega(a)|[\mathds 1^{(\beta)}_{\mathcal A}]\rangle = 0,
\end{equation}
the density matrix $\rho_\omega$ can be rewritten as
\begin{equation}
\rho_\omega = \sum_\alpha | [\mathds 1^{(\alpha)}_{\mathcal A}]\rangle \langle [\mathds 1^{(\alpha)}_{\mathcal A}]|.
\end{equation}
If  we set
\begin{equation}
\lambda_\alpha = \langle[\mathds 1^{(\alpha)}_{\mathcal A}]|[\mathds 1^{(\alpha)}_{\mathcal A}]\rangle,
\end{equation}
we can rewrite $\rho_\omega$ in terms of normalised vectors, as follows. We define
\begin{eqnarray}
|\chi^{(\alpha)}\rangle= \frac{1}{\sqrt{\lambda_\alpha}} |[\mathds 1^{(\alpha)}_{\mathcal A}]\rangle.
\end{eqnarray}
Then,
\begin{equation}
\langle \chi^{(\alpha)} | \chi^{(\beta)}\rangle= \delta_{\alpha\beta}.
\end{equation}
With
\begin{equation}
\rho_\omega^{(\alpha)}:= |\chi^{(\alpha)} \rangle \langle \chi^{(\alpha)} |,
\end{equation}
we then obtain the decomposition of $\rho_\omega$ in terms of pure states as
\begin{equation}
\label{eq:rho-decomposed}
\rho_\omega = \sum_\alpha\lambda_\alpha \rho_\omega^{(\alpha)},
\end{equation}
where
\begin{equation}
\lambda_\alpha>0,\;\;\;\sum_\alpha\lambda_\alpha=1,\;\;\; \rho_\omega^{(\alpha)} \rho_\omega^{(\beta)}= \delta_{\alpha \beta} \rho_\omega^{(\alpha)}.
\end{equation}

The von Neumann entropy of $\rho_\omega$ is
\begin{equation}
S(\rho_\omega)=- \Tr\rho_\omega\log\rho_\omega = -\sum_\alpha \lambda_\alpha\log \lambda_\alpha.
\end{equation}
Corresponding to (\ref{eq:rho-decomposed}), we have the decomposition of $\omega$ into extremal or pure states $\omega^{(\alpha)}$ (Recall that $\omega(\cdot)=\Tr~\rho\cdot $):
\begin{equation}
\omega = \sum_\alpha \lambda_\alpha \omega^{(\alpha)}.
\end{equation}
It has entropy
\begin{equation}
S(\omega)= -\sum_\alpha \lambda_\alpha\log \lambda_\alpha.
\end{equation}

There are important issues related to the uniqueness of the decomposition (\ref{eq:rho-decomposed}) and hence of the entropy of $\omega$ as observed by R. Sorkin
\cite{*[{Private communication.}] [{ }] Sorkin2012}
[\emph{cf}. eq. (\ref{eq:basis-change})]. For a detailed discussion of these issues we refer to \cite{Balachandran2012c}.

\section{On Entanglement and Subalgebras}
\label{sec:3}

For a bipartite system of non-identical particles $A$ and $B$ with Hilbert spaces $\mathcal H_A$ and $\mathcal H_B$, a vector state $|\psi\rangle\in \mathcal H=\mathcal H_A\otimes\mathcal H_B$, which in general is of the form
\begin{equation}
|\psi\rangle = \sum_{i,j}C_{ij}|\chi_{A,i}\rangle \otimes |\eta_{B,j}\rangle,
\end{equation}
is said to be entangled if it cannot be reduced to the form
\begin{equation}
|\psi\rangle= |\chi'_A\rangle\otimes|\eta'_B\rangle
\end{equation}
by  a change of basis.

A measure of entanglement is the von Neumann entropy of the reduced density matrix
\begin{equation}
\rho_A = \Tr_{\mathcal H_B}|\psi\rangle\langle\psi|.
\end{equation}
We assume that $\langle \psi|\psi\rangle=1$. The vector $|\psi\rangle$ is entangled if and only if
\begin{equation}
S(\rho_A)= -\rho_A \log \rho_A \neq 0.
\end{equation}

The physical meaning of the partial trace and the reduced density matrix are as follows.

Suppose that we observe only the subalgebra
\begin{equation}
\mathcal A_0= \lbrace \alpha_0\in \mathcal A\;|\;  \alpha_0= K_A\otimes \mathds 1_B \rbrace,
\end{equation}
where $K_A$ is an observable acting on $\mathcal H_{\mathcal A}$ and $\mathds 1_B$ is the identity operator on $\mathcal H_B$. The algebra $\mathcal A$ of all observables on $\mathcal H=\mathcal H_A\otimes \mathcal H_B$ contains $\mathcal A_0$ as a subalgebra:
\begin{equation}
\mathcal A_0 \subset \mathcal A.
\end{equation}
Now, for restricted observations of just $\mathcal A_0$,
\begin{equation}
\Tr_{\mathcal H} (\rho~\alpha_0) = \Tr_{\mathcal H_A} (\rho_A~K_A), \;\;\alpha_0=K_A\otimes \mathds 1_B\in \mathcal A_0.
\end{equation}
Thus, $\rho_A$ is said to be  the \emph{restriction} of $\rho$ to $\mathcal A_0$.

Let $\omega$ be the state for the density matrix $\rho= |\psi\rangle\langle \psi|$:
\begin{equation}
\omega(\alpha) = \Tr (\rho~\alpha), \;\;\;\alpha\in \mathcal A.
\end{equation}
In the same way, let $\omega_A$ be the state for the density matrix $\rho_A$:
\begin{equation}
\omega_A(\alpha_0) = \Tr_{\mathcal H_A} (\rho_A~K_A),\;\;\;\alpha_0= K_A\otimes \mathds 1_B\in \mathcal A_0
\end{equation}
Then $\omega_A$ is said to be  the \emph{restriction} of the state $\omega$ on $\mathcal A$ to $\mathcal A_0$:
\begin{equation}
\omega_A= \omega|_{\mathcal A_0}.
\end{equation}
Thus partial trace in this case maps a density matrix $\rho$ and a state $\omega$ on $\mathcal A$ to their restrictions
$\rho_A,\omega_A$ on $\mathcal A_0$.

But there are many cases where partial trace cannot be interpreted this way and in fact has no physical meaning. A well-discussed example is that of identical fermions~\cite{Ghirardi2004,Schliemann2001,Banuls2009,Plastino2009}. Denoting antisymmetrisation by $\wedge$,
\begin{equation}
|\psi\rangle\wedge|\chi\rangle=\frac{1}{\sqrt 2}\left(|\psi\rangle\otimes |\chi\rangle- |\chi\rangle\otimes |\psi\rangle\right),
\end{equation}
a generic $N$-particle vector of identical fermions is a linear combination of vectors of the form
\begin{equation}
|\psi\rangle=|\psi_1\rangle\wedge|\psi_2\rangle\wedge\ldots\wedge|\psi_N\rangle.
\end{equation}
It lives in the $N$-fold  antisymmetric product $\mathcal H$ of the one-particle Hilbert space $\mathcal H^{(1)}$:
\begin{equation}
\mathcal H= \mbox{$\bigwedge^{N}$} \mathcal H^{(1)}, \;\;\;|\psi\rangle\in \mathcal H.
\end{equation}
The algebra $\mathcal A$ of \emph{observables} must necessarily leave $\mathcal H$ invariant. That means that observables must be permutation invariant. An operator  such as $K_1\otimes \mathds 1\otimes \cdots\otimes \mathds 1$
($\mathds 1$ is the identity on $\mathcal H^{(1)}$ and $K_1\neq \mathds 1$) is not permutation invariant and, therefore, is not an observable.
Hence now \emph{partial traces do not correspond to restrictions to subalgebras of observables on}
 $\mathcal H$.

 So, in a generic problem, partial trace is not a meaningful operation.

 But the restriction of a state $\omega$ on $\mathcal A$ to a subalgebra $\mathcal A_0$ is always sensible. What we need is a criterion to select $\mathcal A_0$ appropriately for a physical question.

 For example, the algebra $\mathcal A_0$ appropriate for single particle observables is generated by
 \begin{equation}
 \label{eq:K-times-K}
 K\otimes \mathds 1\otimes\cdots\otimes \mathds 1\; + \;  \mathds 1\otimes K\otimes\cdots\otimes \mathds 1
 \; + \; \cdots+ \mathds 1\otimes \cdots\otimes\mathds 1\otimes K,
\end{equation}
where $K$ is an observable on $\mathcal H^{(1)}$.

Choices such as (\ref{eq:K-times-K}) are dictated by a coproduct on the single-particle algebra. We will return to this point later.

For such reasons, as declared in the Introduction, \emph{entanglement is a property characterising  a triple}  $(\mathcal A, \mathcal A_0,\omega)$. \emph{We avoid the use of partial trace}.

\section{Examples}
\label{sec:4}

\subsection{The Algebra $M_2(\mathds C)$}
\label{sec:2x2-matrix-example}
 The choice $\mathcal A= M_2(\mathds C)$ of $2\times 2$ complex matrices is a simple non-trivial
 example to illustrate the GNS construction. It is discussed already in appendix 3 of Landi~\cite{Landi2008}. We will recall
 this example here and, in addition,  use it to illustrate the entropy calculation.

 The algebra $\mathcal A$ acts on $\mathds C^2$. Let
 \begin{equation}
 \lbrace |i\rangle:\; i=1,2,\;\;\langle i | j \rangle = \delta_{ij}  \rbrace
 \end{equation}
 be an orthonormal basis of $\mathds C^2$. Then the matrix units
 \begin{equation}
 e_{ij} = | i \rangle\langle j|
 \end{equation}
 span $M_2(\mathds C)$. Note that
 \begin{equation}
 \label{eq:e_ij}
 e_{ij}e_{kl}=\delta_{jk} e_{il}.
 \end{equation}
 An element $\alpha$ of $\mathcal A$ can be expanded in this basis as
 \begin{equation}
 \alpha = \sum_{i,j} \alpha_{ij} e_{ij}.
 \end{equation}
 Following \cite{balachandran2012a},    for the state $\omega$ we choose
 \begin{equation}
 \label{eq:state_2x2}
 \omega(\alpha) = \lambda \alpha_{11} + (1-\lambda) \alpha_{22},\;\;\; 0 \leq \lambda \leq 1.
 \end{equation}
 The null space $\mathcal N_\omega$ is determined  by the condition
 \begin{equation}
 \omega(\alpha^*\alpha)=0.
 \end{equation}
 For our choice (\ref{eq:state_2x2}) for $\omega$ we obtain, making use of (\ref{eq:e_ij}),
 \begin{equation}
 \omega(\alpha^* \alpha)= \lambda (|\alpha_{11}|^ 2+|\alpha_{21}|^2) +
 (1-\lambda) (|\alpha_{12}|^2+|\alpha_{22}|^2).
 \end{equation}
 The solution depends on $\lambda$. We consider 3 cases.

 \emph{\bf Case 1}: $\lambda=0$.

 In this case, $\alpha\in \mathcal N_\omega$ if $\alpha_{12}=\alpha_{22}=0$. So
 \begin{equation}
 \mathcal N_\omega=\left\lbrace\left( \begin{array}{cc}
                                                    \alpha_{11} & 0\\
                                                    \alpha_{21} & 0
                                                   \end{array}\right)\;: \;\alpha_{11},\alpha_{21}\in \mathds C\right\rbrace \cong\mathds C^2.
 \end{equation}
Since $\hat{\mathcal A}\cong \mathds C^4$, we obtain
 \begin{eqnarray}
 \mathcal H_\omega= \hat{\mathcal A} /\hat{\mathcal N_\omega} \cong \mathds C^2,
 \end{eqnarray}
 with basis
 \begin{equation}
 \label{eq:2x2-basis}
\left\lbrace |[e_{k2}]\rangle\right\rbrace_{k=1,2}.
 \end{equation}

 The representation $\pi_\omega$ of $\mathcal A$ on $\mathcal H_\omega$ is given by
 \begin{equation}
 \pi_\omega(e_{ij})|[e_{k2}]\rangle = \delta_{jk} |[e_{i2}]\rangle.
 \end{equation}
 It is isomorphic to the defining representation on $\mathds C^2$. Therefore it is irreducible. So we conclude that
 $\rho_\omega$ is a rank 1 projector and  has vanishing entropy:
 \begin{equation}
 S(\rho_\omega)=0.
 \end{equation}

 \emph{\bf Case 2}: $\lambda=1$.

 This is similar to the case $\lambda =0$. The null space $\hat{\mathcal N_\omega}\cong\mathds C^2$ is spanned by
 \begin{equation}
 |e_{12}\rangle,\;\;|e_{22}\rangle.
 \end{equation}
 and  $\mathcal H_\omega= \hat{\mathcal A} /\hat{\mathcal N_\omega}$ has basis
 \begin{equation}
 \label{eq:2x2-basisB}
\left\lbrace |[e_{k1}]\rangle\right\rbrace_{k=1,2}.
 \end{equation}
 The representation $\pi_\omega$ is irreducible, isomorphic to the $\lambda=0$ representation and carries zero entropy.

\emph{\bf Case 3}: $0<\lambda<1$.

There are no non-zero null vectors in this case:
\begin{equation}
\mathcal N_{\omega_\lambda}= \lbrace 0 \rbrace.
\end{equation}
Hence
\begin{eqnarray}
 \mathcal H_\omega= \hat{\mathcal A} /\hat{\mathcal N_\omega} \cong \mathds C^4,
 \end{eqnarray}
 It has basis
 \begin{equation}
 \lbrace |[e_{ij}]\rangle\rbrace_{i,j=1,2}.
 \end{equation}
 The representation $\pi_\omega$ is given by
 \begin{equation}
 \pi_\omega(e_{ij})|[e_{kl}] \rangle=|[e_{ij} e_{kl}] \rangle= \delta_{jk}|[e_{il}]\rangle
 \end{equation}
 This representation is reducible into two two-dimensional irreducible ones:
 $\mathcal H_\omega=\mathds C^2\oplus \mathds C^2$.
 The first $\mathds C^2$ has basis
 \begin{equation}
 \label{eq:first-C}
 \lbrace e_{a 1}\rbrace_{a=1,2}
 \end{equation}
and the second
 \begin{equation}
 \label{eq:second-C}
 \lbrace e_{a 2}\rbrace_{a=1,2}
 \end{equation}
 We must next express $|[\mathds 1_{\mathcal A}]\rangle$ in terms of its components in these subspaces. That is easy:
 \begin{equation}
 |[\mathds 1_{\mathcal A}]\rangle= |[e_{11}]\rangle + |[e_{22}]\rangle.
 \end{equation}
 It follows that $\omega_\lambda$ is not pure and can be expressed in terms of the following density matrix:
 \begin{equation}
 \rho_{\omega_\lambda}= |[e_{11}]\rangle\langle[e_{11}]| + |[e_{22}]\rangle\langle[e_{22}]|.
 \end{equation}
 Since
 \begin{eqnarray}
 \langle [e_{11}] | [e_{11}]\rangle & = & \omega (e_{11}^* e_{11})= \omega (e_{11}) = \lambda\nonumber\\
 \langle [e_{22}] | [e_{22}]\rangle & = & \omega (e_{22}^* e_{22})= \omega (e_{22}) = 1-\lambda,
 \end{eqnarray}
 \begin{equation}
 \rho_{\omega_\lambda} = \lambda \rho_{11} +(1-\lambda)\rho_{22},
 \end{equation}
 where $\rho_{11}$ and $\rho_{22}$ are the rank 1 density matrices
 \begin{eqnarray}
 \rho_{11} &=& \frac{1}{\lambda} |[e_{11}]\rangle\langle[e_{11}]|,\nonumber\\
 \rho_{22} &=& \frac{1}{1-\lambda} |[e_{22}]\rangle\langle[e_{22}]|.
 \end{eqnarray}
 We can read off the entropy of $\rho_{\omega_\lambda}$ to be
 \begin{equation}
 \label{eq:4.26}
 S(\rho_{\omega_\lambda})=-\lambda \log \lambda -(1-\lambda) \log (1-\lambda).
 \end{equation}

 \emph{Remark}:

 This example shows that the irreducible representations of dimension 2 occur  with multiplicity 2.  This is a general feature: for any representation $\pi$ of a $C^*$-algebra $\mathcal A$ with a \emph{cyclic and separating
 vector}\footnote{A vector state $|\Omega\rangle\in\mathcal{H}$ is cyclic when $\pi(\mathcal{A})|\Omega\rangle$ is dense in $\mathcal{H}$. A state $|\Omega\rangle\in\mathcal{H}$ is separating when the map $\mathcal{A}\to \pi(\mathcal{A})|\Omega\rangle$  is injective.}~\cite{haag1996local}, an irreducible representation of dimension $d$ occurs with multiplicity $d$. This is as in regular representations of compact groups and follows from Tomita-Takesaki theory~\cite{haag1996local}

 Rafael Sorkin has pointed out to us \cite{*[{Private communication.}] [{ }] Sorkin2012} that if  $d>1$, the splitting of $\mathcal{H}_\omega$ into irreducible subspaces such as $\mathcal{H}=\mathbb{C}^2\oplus\mathbb{C}^2$ before (\ref{eq:first-C}) is not unique. For example, we could have chosen another pair of $\mathds C^2$'s with basis
\begin{align}
\label{eq:basis-change}
\lefteqn{
\sum_{\alpha}\xi_\alpha |[e_{a\alpha}]\rangle,\;\;\;  \sum_{\alpha}\eta_\alpha |[e_{a\alpha}]\rangle,
              { } }\nonumber\\
 &  { }  \xi^\dagger\cdot \xi,\;\;\eta^\dagger\cdot \eta\neq 0,\;\;\xi^\dagger\cdot\eta=0 { }
 \end{align}
and recalculated  $\rho_{\omega_\lambda}$  and its entropy. They depend on $\xi,\eta$. This feature is generic
for $d>1$. The entropy (\ref{eq:4.26}) is the least one. Further discussion of such issues will be given elsewhere \cite{Balachandran2012c}.

\subsection{A $\mathds C^2\otimes \mathds C^2$ example}

The matrices
\begin{equation}
\sigma_\mu:\;\;\;\sigma_0=\mathds 1_2,\;\;\sigma_{i}= \mbox{Pauli matrices},
\end{equation}
form a basis for $M_2(\mathds C)$. So the algebra $\mathcal A= M_4(\mathds C)$ on $\mathds C^2\otimes \mathds C^2$
is generated by
\begin{equation}
\sigma_\mu\otimes\mathds 1_2,\;\;\; \mathds 1_2\otimes\sigma_\mu,\;\;\;\mu=0,\ldots,3.
\end{equation}

For the state $\omega\equiv \omega_\theta$, let us choose :
\begin{equation}
\rho_{\omega_\theta}=|\psi_\theta\rangle\langle\psi_\theta |,\;\;\;|\psi_\theta\rangle=\cos\theta |+-\rangle - \sin\theta|-+\rangle.
\end{equation}
Here
\begin{equation}
|+-\rangle = |+\rangle\otimes|-\rangle, \;\;\;\;\textrm{etc.}
\end{equation}
 with
 \[
 |+\rangle= \left(\begin{array}{c} 1\\0\end{array} \right), \;\;\;\; |-\rangle= \left(\begin{array}{c} 0\\1\end{array} \right).
 \]
Let us now choose, for the subalgebra $\mathcal A_0$, the subalgebra generated by the (``local'') operators
\begin{equation}
\lbrace \sigma_\mu\otimes \mathds 1_2\rbrace_{\mu=0,\ldots,3}.
\end{equation}
In this context, entanglement can be understood in terms of correlations between measurements performed by two observers $A$ and $B$ having access to observables corresponding only to  $\mathcal A_0$ (in the case of, say,  $A$) and observables corresponding only to the commutant of $\mathcal A_0$ (in the case of $B$).

For  $A$, the state $\omega_\theta$ becomes the restriction of $\omega_\theta$ to $\mathcal A_0$:
\begin{equation}
\omega_{\theta,0}=\omega_\theta|_{\mathcal A_0}.
\end{equation}
Notice that, \emph{in this case}, the result of the restriction coincides with the one obtained by partial trace. In fact, since every element of $\mathcal A_0$ is of the form  $\tilde \alpha=\alpha\otimes \mathds 1_2$ (for some $\alpha\in M_2(\mathds C)$), we obtain:
\begin{eqnarray}
\label{eq:qubits}
\omega_{\theta,0}(\tilde \alpha) &=& \langle \psi_\theta |(\alpha\otimes \mathds 1_2)|\psi_\theta\rangle\nonumber\\
&=& \cos^2\theta\langle+|\alpha|+\rangle + \sin^2\theta\langle-|\alpha|-\rangle\\
&=& \cos^2\theta\; \alpha_{11} + \sin^2\theta \;\alpha_{22} \nonumber
\end{eqnarray}
Taking into account the fact that $\mathcal A_0\cong M_2(\mathds C)$, and comparing equations (\ref{eq:qubits}) and (\ref{eq:state_2x2}), we see that the entropy obtained from the GNS construction is given by (\ref{eq:4.26}) upon replacing $\lambda$ by $\cos^2\theta$:
\begin{equation}\label{eq:4.S}
S(\theta)= -\cos^2\theta\log\cos^2\theta -\sin^2\theta\log\sin^2\theta.
\end{equation}
Clearly, $S(\theta)$ corresponds to the entanglement of the vector state $|\psi_\theta\rangle$. In particular, for $\theta=\pi/4$ the state is the (maximally entangled) Bell vector state
\begin{equation}
\label{eq:4.Bell}
|\psi_{\theta=\frac{\pi}{4}}\rangle=\frac{1}{\sqrt 2}\left(|+-\rangle - |-+\rangle\right),
\end{equation}
for which (\ref{eq:4.S}) reduces to $\log 2$.

Now, in order to illustrate how to deal with cases where $\mathcal A_0$ does not act on just one factor of a bipartite system, we consider the  Bell state (\ref{eq:4.Bell}) together with different choices for $\mathcal A_0$. We focus on choices for which partial trace has no meaning. For instance, we now consider the following three choices:
\begin{itemize}
\item $\mathcal A_\pm$, generated by $\lbrace\sigma_\mu\otimes\left(\frac{1\pm\sigma_3}{2}\right)\rbrace_\mu$.
\item $\mathcal A_+\oplus \mathcal A_-$.
\end{itemize}
(Note that $\mathcal A_+ \mathcal A_-=\lbrace 0\rbrace$).

\emph{\bf Case 1}: $\mathcal A_0=\mathcal A_+$.

The null space $\widehat{\mathcal N}_\omega^+\subset \hat{\mathcal A}_+$ is determined by the equation
\begin{equation}
\rho_\omega(\alpha^* \alpha)=\langle \psi_{\theta=\frac{\pi}{4}}|\alpha^*\alpha | \psi_{\theta=\frac{\pi}{4}}\rangle=0,\quad \textrm{for}\;\;\alpha \in \mathcal A_+.
\end{equation}
or
\begin{equation}
\alpha|\psi_{\theta=\frac{\pi}{4}}\rangle = 0.
\end{equation}
Hence
\begin{equation}
\widehat{\mathcal N}_\omega^+ =
 \left\lbrace    \left|   \left(\begin{array}{cc} \alpha_{11} & 0\\
                                                            \alpha_{21} & 0\end{array}     \right)
                                                            \otimes\left(\frac{1+\sigma_3}{2}\right) \right\rangle
                                                            \;\;:\;\;\alpha_{i1}\in \mathds C   \right\rbrace \cong \mathds C^2.
\end{equation}
The quotient space $\hat{\mathcal A} / \widehat{\mathcal N}_\omega^+\cong \mathds C^2$ is spanned
by
\begin{equation}
 \left\lbrace    \left| \left[  \left(\begin{array}{cc}    0 & \alpha_{12}\\
                                                                       0 & \alpha_{22}\end{array}     \right)
                                                            \otimes\left(\frac{1+\sigma_3}{2}\right) \right]\right\rangle
                                                            \;\;:\;\;\alpha_{i2}\in \mathds C   \right\rbrace.
\end{equation}
It transforms irreducibly under $\mathcal A^+$. So $\rho_\omega$ remains pure with zero entropy.

\emph{\bf Case 2}: $\mathcal A_0=\mathcal A_-$.

This is similar to case 1. The null space is

\begin{equation}
\widehat{\mathcal N}_\omega^-=
 \left\lbrace    \left|   \left(\begin{array}{cc}    0 & \alpha_{12}\\
                                                                       0 & \alpha_{22}\end{array}     \right)
                                                            \otimes\left(\frac{1-\sigma_3}{2}\right) \right\rangle
                                                            \;\;:\;\;\alpha_{i2}\in \mathds C   \right\rbrace\cong \mathds C^2,
\end{equation}
the quotient space being
\begin{equation}
\hat{\mathcal A}^-/\widehat{\mathcal N}_\omega^-=
 \left\lbrace    \left|\left[   \left(\begin{array}{cc} \alpha_{11} & 0\\
                                                                            \alpha_{21} & 0\end{array}     \right)
                                                            \otimes\left(\frac{1-\sigma_3}{2}\right)\right] \right\rangle
                                                            \;\;:\;\;\alpha_{i1}\in \mathds C   \right\rbrace \cong \mathds C^2.
\end{equation}

It transforms irreducibly under $\mathcal A^-$. So $\rho_\omega$ remains pure with zero entropy.

\emph{\bf Case 3}: $\mathcal A_0=\mathcal A_+\oplus \mathcal A_-$

The null space $\widehat{\mathcal N}_\omega$ is the direct sum
\begin{equation}
\widehat{\mathcal N}_\omega=\widehat{\mathcal N}_\omega^+\oplus\widehat{\mathcal N}_\omega^-,
\end{equation}
while the quotient space is
\begin{equation}
\label{bell-case-3}
(\hat{\mathcal A}_+\oplus \hat{\mathcal A}_-)/\widehat{\mathcal N}_\omega =
 \left\lbrace    \left|\left[
                                                      \left(\begin{array}{cc}    0 & \alpha_{12}\\
                                                                                          0 & \alpha_{22}\end{array}     \right)
                                                            \otimes\left(\frac{1+\sigma_3}{2}\right)+
 \left(\begin{array}{cc} \alpha_{11} & 0\\
                                    \alpha_{21} & 0\end{array}     \right)
                                                            \otimes\left(\frac{1-\sigma_3}{2}\right)\right] \right\rangle
                                                            \right                            \rbrace.
\end{equation}
This representation is the direct sum of two irreducible representations given by the two terms in (\ref{bell-case-3}). We must now restrict
\begin{equation}
|\mathds 1_{\mathcal A}\rangle= \left(\begin{array}{cc} 1 & 0\\0 & 1\end{array}\right) \otimes \left(\begin{array}{cc} 1 & 0\\0 & 1\end{array}\right)
\end{equation}
into its components in  $(\hat{\mathcal A}_+\oplus \hat{\mathcal A}_-)/\widehat{\mathcal N}_\omega$.

Its component in $\hat{\mathcal A}_+/\widehat{\mathcal N}_\omega^+$ is
\begin{equation}
\left(\begin{array}{cc} 0 & 0\\0 & 1\end{array}\right) \otimes \left(\begin{array}{cc} 1 & 0\\0 & 0\end{array}\right) := \mathds 1^+,
\end{equation}
while that in $\hat{\mathcal A}_-/\widehat{\mathcal N}_\omega^-$ is
\begin{equation}
\left(\begin{array}{cc} 1 & 0\\0 & 0\end{array}\right) \otimes \left(\begin{array}{cc} 0 & 0\\0 & 1\end{array}\right) := \mathds 1^-.
\end{equation}
The squared norm of $|[\mathds 1^\pm]\rangle$ is $1/2$:
\begin{equation}
\langle[\mathds 1^\pm] | [\mathds 1^\pm] \rangle =  \Tr\left( \rho_\omega (\mathds 1^\pm)^* \mathds 1^\pm \right)
  =  \frac{1}{2}.
 \end{equation}
 Hence
 \begin{equation}
 \rho_\omega|_{(\mathcal A_+\oplus \mathcal A_-)}=
 \frac{1}{2}\left(
 \sqrt 2 |[\mathds 1^+]\rangle\langle[\mathds 1^+]|\sqrt 2 + \sqrt 2 |[\mathds 1^-]\rangle\langle[\mathds 1^-]|\sqrt 2
 \right),
 \end{equation}
 giving $S(\rho_\omega|_{(\mathcal A_+\oplus \mathcal A_-)})=\log 2$ for the entropy.

 \subsection{Role of Hopf algebras}

In elementary quantum physics, one starts with a Hilbert space $\mathcal{H}_{A_i}$ which typically carries the representation of the algebra $\mathcal{A}_{A_i}$ of single-particle observables for particle $A_i$. In the second quantized version, there is an isomorphism of  $\mathcal{A}_{A_i}$ into the full algebra $\mathcal{A}_{A_1,A_2,\ldots,A_k}$  of observables on
\begin{equation}
\mathcal{H}_{A_1,A_2,\ldots,A_k} =\mathcal{H}_{A_1}\otimes\cdots\otimes\mathcal{H}_{A_k},
\end{equation}
if the particles $A_1,A_2,\ldots,A_k$ are non-identical.

We must identify the $A_1$-particle observables in the $k$-particle
Hilbert space to be able to observe properties of $A_1$ say in $\mathcal{H}_{A_1,A_2,\ldots,A_k}$. They are given by the isomorphism $\Delta^k$ for non-identical particles defined by
\begin{equation}
\label{eq:Delta_k-dist}
\Delta^k(\alpha_{A_1})=\alpha_{A_1}\otimes \mathds 1_{A_2}\otimes\cdots\otimes \mathds 1_{A_k}, \;\;\;\alpha_{A_1}\in\mathcal{A}_{A_1},
\end{equation}
as we saw earlier.

When the particles are identical so that $A_1=A_2=\cdots=A_k$ and are fermions or bosons, such an isomorphism still exists: they are the totally symmetrized versions of (\ref{eq:Delta_k-dist}) as we also saw (cf. (\ref{eq:K-times-K})). This is, in fact, the simplest choice of a coproduct
\begin{equation}
  \Delta\equiv \Delta^{1},
\end{equation}
given by
\begin{equation}
\label{eq:coproduct-simplest}
\Delta^k(\alpha) := \alpha\otimes\mathds 1\otimes\cdots \otimes \mathds 1\;\; +\;\;
\mathds 1\otimes \alpha \otimes\cdots \otimes \mathds 1\;\;  + \; \cdots \;  +\;\;\mathds 1\otimes\cdots \otimes\alpha.
\end{equation}

When the particles are identical, but fulfill braid group statistics, there still exists $\Delta^k$ as we will show below. As a special case, the same holds for parastatistics, its expression then is the same as for bosons and fermions.

The  importance of $\Delta^k$ is as follows: \emph{it identifies the single particle observables in the $k$-particle Hilbert space}.

The choice of $\Delta^k$ typically comes from the statistics group. It can differ if the latter differs. If the statistics group is the braid group or its quotient, the permutation group, then $(\mathcal{A}_{A_i}, \Delta^k)$ defines a ``quasi-triangular Hopf algebra''~\cite{Balachandran2010}. The isomorphism $\Delta^1 \equiv \Delta$ then defines a ``coproduct''  from which $\Delta^k$ can be deduced.

More general possibilities than braid group and accordingly more general single-particle algebras than the above can also be contemplated \cite{Aneziris1989, Aneziris1991,Balachandran2011}.

There is conceptually no problem in restricting a state $\omega$ on $\mathcal{A}_{A_1,A_2,\ldots,A_k}$ to the subalgebra $\Delta^k(\mathcal{A}_{A_i})$ and comparing the entropies of $\omega|_{\mathcal{A}_{A_1,A_2,\ldots,A_k}}$ and
$\omega|_{\Delta^k(\mathcal{A}_{A_i})}$.

We may also wish to study a group of $k$-particles with an algebra $\mathcal{A}^{(k)}$ in an $n$-particle Hilbert space, for any $n>k$. If $\mathcal A^{(k)}$ is Hopf, then we can find its isomorphic algebra at the $n$-particle level, often more than one. The ambiguity  in its choice has to be resolved  by the context. Entropy considerations  can again be pushed through.

\subsection{Identical Particles}
\label{sec:identical-particles}

 In this section we illustrate the use of the GNS construction for the evaluation of entanglement entropy in systems of
 identical particles, making use of the coproduct in order to identify subalgebras of one-particle observables. The
 general setting for the three examples we consider below is the following. We consider a one-particle Hilbert space $\mathcal
 H^{(1)}\cong
 \mathds C^d$. In this case, the full one-particle observable algebra is given by the group algebra of $U(d)$,
 $\mathds C U(d)$. The two-particle Hilbert space is then given by the subspace of $\mathcal H^{(1)}\otimes \mathcal
 H^{(1)}$ consisting of either \emph{symmetric}  (bosonic statistics) or \emph{antisymmetric} (fermionic statistics) tensors. The coproduct is a homomorphism
 $\Delta: \mathds C U(d) \rightarrow \mathds C U(d)\otimes \mathds C U(d)$ that allows us to map one-particle
 observables from the one-particle sector to the two-particle sector. The map $\Delta$ is not fixed \emph{a priori}.
 Here we consider the standard choice
 \begin{equation}
 \label{eq:standard-coproduct}
\Delta (g) = g\otimes g,
 \end{equation}
 for $g\in U(d)$, linearly extended to all of $\mathds C U(d)$. This coproduct is the exponentiated form of \ref{eq:coproduct-simplest} (see also below). The crucial property is \emph{coassociativity}:
 \begin{equation}
 \label{eq:coassociativity}
(\Delta \otimes \mbox{id}) \Delta = ( \mbox{id} \otimes \Delta ) \Delta.
\end{equation}
This property allows us, starting from $\mathds C U(d)$ and via the coproduct, to construct observables acting at the
$k$-particle level, for any $k$. In the next section, other choices of the coproduct will be used in order to apply our
ideas to examples with braid group statistics.

We always consider algebras $\mathcal A$ with unity and subalgebras $\mathcal A_0$ which contain this unity. The physical reason for including the unity of $\mathcal A$ in $\mathcal A_0$ will become apparent in section \ref{sec:6}.

Now, if we perform measurements where only a restricted set of one-particle observables is considered, we may study the
entanglement of a given two-particle state $|\psi\rangle$ that arises from the corresponding restriction. For example,
if from the $d$ available ``levels'' in $\mathcal H^{(1)} = \mathds C^d$ we consider only $d\,'$ of them ($d\,'< d$),
the algebra of one-particle observables will be reduced to the algebra generated by  $\mathds C U(d\,')$ and unity. Its dimension is $d\,'^{\,2} +1$. The two-particle Hilbert space will
then decompose into irreducible representations of this algebra, this being directly reflected in the
entanglement structure of the state $|\psi\rangle$.

 \subsubsection{{\bf Two Fermions, $\mathcal{H}^{(1)}=\mathds{C}^4$}}

 As a first example, we consider the case $d=4$, $d\,'=2$. The two-fermion space $\Lambda^2 \mathcal H^{(1)}$ is
6-dimensional, as can be seen from the decomposition $4\otimes 4 = 10 \oplus 6$ of $\mathds{C}^4\otimes \mathds{C}^4$
into symmetric and antisymmetric tensors. Let us consider a particular orthonormal basis
$\lbrace|e_1\rangle,|e_2\rangle,|e_3\rangle,|e_4\rangle\rbrace$ for $\mathcal H^{(1)}$. A basis for the two-fermion
space $\Lambda^2 \mathcal H^{(1)}\cong \mathds C^6$ is then given by $\lbrace|e_i\rangle \wedge |e_j\rangle
\rbrace_{1\leq i<j\leq 4}$.

Now we will assume that \emph{only} one-particle observables containing  $\mathds 1_4$ and those causing transitions between  the states $|e_1\rangle$ and $|e_2\rangle$ are considered. Then, the relevant algebra of observables is isomorphic to $\mathds C U(2)\otimes \mathds 1_4$. Using
the coproduct, we can obtain the image of this algebra  acting on the two-particle sector. For our choice
of the subalgebra, the relevant observables are generated by operators of the form $M_{ij}=
|e_i\rangle \langle e_j|$, with $1\leq i,j \leq 2$ and $\mathds 1_4$.  In this context, it turns out to be useful to  work with the
``infinitesimal'' version of (\ref{eq:standard-coproduct}), that is, if $L$ is an element of the \emph{Lie algebra} of
$U(d)$, we set
\begin{equation}
\label{eq:coproduct-infinitesimal} \Delta(L)= L\otimes\mathds 1 + \mathds 1 \otimes L.
\end{equation}
%
It is also convenient to label the basis vectors of $\Lambda^2 \mathcal H^{(1)}$ in the following way:
\begin{eqnarray}
& |a\rangle = | e_1\rangle \wedge | e_2\rangle\equiv \frac{1}{\sqrt{2}}\left(| e_1\rangle \otimes | e_2\rangle-| e_2\rangle \otimes | e_1\rangle \right), &\\
& | \alpha_1\rangle=| e_1\rangle \wedge | e_3 \rangle,\;\; | \alpha_2\rangle=|e_2 \rangle \wedge | e_3\rangle &\\
& | \beta_1\rangle=| e_1\rangle \wedge | e_4 \rangle,\;\; | \beta_2\rangle=|e_2 \rangle \wedge | e_4\rangle &\\
& |b \rangle=|e_3 \rangle \wedge | e_4\rangle &
\end{eqnarray}
From  (\ref{eq:coproduct-infinitesimal}) it is easy to obtain explicit expressions for the matrix representations of the relevant one-particle observables. As an
illustration, we compute:
\begin{eqnarray}
\Delta (M_{12})|\alpha_2\rangle & =& \Delta (M_{12}) |e_2 \rangle \wedge | e_3\rangle \nonumber\\
 &= & (M_{12} \otimes \mathds 1  + \mathds 1\otimes M_{12}) \frac{1}{\sqrt{2}}(|e_2 \rangle \otimes | e_3\rangle - |e_3 \rangle \otimes |
 e_2\rangle)\nonumber\\
 &=& \frac{1}{\sqrt{2}}(|e_1 \rangle \otimes | e_3\rangle - |e_3 \rangle \otimes |
 e_1\rangle)\nonumber\\
 &=& |\alpha_1\rangle.
\end{eqnarray}
The four matrices $A_{ij}\equiv \Delta (M_{ij})$ (for $i,j= 1,2$)  turn out to be block diagonal in the chosen basis:
\begin{eqnarray}
\label{eq:Aij}
A_{11} &=& \mbox{diag}\lbrace 1, e_{11}, e_{11}, 0\rbrace,\nonumber\\
A_{22} &=& \mbox{diag}\lbrace 1, e_{22}, e_{22}, 0\rbrace,\nonumber\\
A_{12} &=& \mbox{diag}\lbrace 0, e_{12}, e_{12}, 0\rbrace,\nonumber\\
A_{21} &=& \mbox{diag}\lbrace 0, e_{21}, e_{21}, 0\rbrace,
\end{eqnarray}
where $e_{ij}$ denote the standard matrix units on $M_2(\mathds C)$, i.e., $e_{11}= \left(\begin{array}{cc} 1 & 0\\
0& 0\end{array}\right)$, and so on. To this must be added the unit matrix $\mathds 1_6$.

 The algebra $\mathcal A_0$ will be generated by exponentials of
these matrices and their products.
It has a basis consisting of the five matrices $A_{11},A_{22}, A_{12}, A_{21}$ and $\mathds 1_6$.
This same example can be worked using creation/annihilation operators. In that case, as explained in \cite{balachandran2012a}, the algebra corresponding to observables involving only the identity and transitions between $|e_1\rangle$ and
$|e_2\rangle$ is six dimensional. However, one of the basis elements corresponds to a 2-particle observable, being the product of the number operators of particles 1 and 2. The 1-particle observable algebra is therefore the one obtained after taking the quotient by the ideal generated by that basis element. Here we are using coproducts in such and only 1-particle observables appear upon application of the homomorphism $\Delta$.

We now consider a $\theta$-dependent state vector, given by
\begin{equation}
|\psi_\theta\rangle=\cos\theta |\beta_1\rangle + \sin\theta |\alpha_2\rangle. \label{thetastate}
\end{equation}
As mentioned above, at the two-particle level the full observable algebra $\mathcal A$ given by $M_{6}(\mathds C)$. The
subalgebra of one-particles we have chosen is the  $\mathcal A_0$ constructed above. We proceed to the construction of
the GNS representation corresponding to different values of $\theta$ when the state $|\psi_\theta\rangle$ is restricted
to $\mathcal A_0$.

\noindent{\emph{\bf Case 1}: $0 < \theta < \frac{\pi}{2}$}

When  $0<\theta < \pi/2$ we can easily check that the only non-zero elements $\alpha\in \mathcal A_0$ for which
\begin{equation}
\omega_\theta (\alpha^*\alpha)\equiv \langle \psi_\theta |\alpha^*\alpha|\psi_\theta\rangle =0,
\end{equation}
are linear combinations of $B=\mbox{diag}\lbrace 1, 0_2, 0_2, 0\rbrace$ and $\mathds 1_6-A_{11}-A_{22}$, that is the null space $\mathcal N_\theta$ is generated by these elements. The
GNS Hilbert space $\mathcal H_\theta$ is thus four-dimensional.

Let $\pi_\theta$ denote the corresponding GNS representation of $\mathcal A_0$ on $\mathcal H_\theta$. A convenient
basis for $\mathcal H_\theta$ is given by $\lbrace|[A_{ij}]\rangle\rbrace_{i,j=1,2}$. A straightforward computation
shows that the subspace spanned by $|[A_{12}]\rangle$ and $|[A_{22}]\rangle$,  as well as the subspace
spanned by $|[A_{11}]\rangle$ and $|[A_{21}]\rangle$, are irreducible. The two representations are isomorphic.

The corresponding  projections $P_1$ and $P_2$ can then be used  in order to obtain the components of $|[\mathds
1_6]\rangle$ in each irreducible subspace. From
\begin{equation}
|[\mathds 1_6]\rangle = |[A_{11} + A_{22}]\rangle
\end{equation}
we obtain
\begin{equation}
P_1|[\mathds 1_6]\rangle= |[A_{11}]\rangle,~~ P_2|[\mathds 1_6]\rangle= |[A_{22}]\rangle.
\end{equation}
Using (\ref{thetastate}), we compute
\begin{equation}
\|P_1|[\mathds 1_6]\rangle\|^2 = \cos^2\theta,~~~~~ \|P_2|[\mathds 1_6]\rangle\|^2 = \sin^2\theta.
\end{equation}
A density matrix acting on the GNS space of the restricted state can be obtained as explained above and its entropy can
be computed. The result is:
\begin{equation}
S(\theta) = -\cos^2 \theta \log \cos^2\theta -\sin^2 \theta \log \sin^2\theta.
\end{equation}

\noindent{\emph{\bf Case 2}: $\theta=0$.}

In this case we have
\begin{equation}
|\psi_0\rangle = |\beta_1\rangle.
\end{equation}
So, the space of null vectors is given by the four-dimensional space
\begin{equation}
\mathcal N_{0,0} = \mbox{Span}\left\{|B\rangle, |[{\mathds 1_{6}} - A_{11}] \rangle, |[A_{22}]\rangle,
|[A_{12}]\rangle\right\}. \label{null}
\end{equation}
This means that $\mathcal H_{\theta=0}\cong \mathds C^2$. Hence, the representation is irreducible so that the
corresponding entropy vanishes.

The situation is completely equivalent for the case $\theta = \frac{\pi}{2}$.

Thus, $\mathcal H_\theta$ decomposes into irreducible subspaces according to the following pattern:
 \begin{eqnarray}
\mathcal H_\theta\cong \left\lbrace\begin{array}{cc}
\mathds C^2,&\;\theta=0, \pi/2\\
\mathds C^4\cong \mathds C^2\oplus \mathds C^2, &\; \theta\in (0,\pi/2).\\
\end{array}
\right.
\end{eqnarray}
The significant aspect of this example is the fact that for the values of $\theta $ for which the
Slater rank of $|\psi_\theta\rangle$ is one, namely $\theta = 0$ and $\frac{\pi}{2}$,
we obtain exactly zero for the entropy. In previous
treatments of entanglement for identical particles, the minimum value for the von Neumann entropy
of the reduced density matrix (obtained by partial trace) has been found to be $\log 2$.
 This has been
a source of embarrassment: it seems to suggest that different entanglement criteria have to be adopted,
depending on whether one is dealing with non-identical particles, or with bosons, or fermions.
For a  critical review of previous attempts at a solution to this problem, see \cite{Tichy2011}.
We have shown here  that, by replacing the notion of \emph{partial trace} by the more general
one of \emph{restriction to a subalgebra}, all cases can be treated on an equal footing.


 \subsubsection{{\bf Two Fermions, $\mathcal{H}^{(1)}=\mathds{C}^3$}}

As a second example, we consider the case $d=3$, with two choices for $d'$: $d'=2, 3$. In this case,  the
two-fermion space is given by $\mathcal H^{(2)}=\Lambda^2 \mathds C^3\subset\mathcal H^{(1)}\otimes \mathcal H^{(1)}$.
Let  $\big\{|e_1\rangle, |e_2\rangle,|e_3\rangle\big\}$ denote an orthonormal basis for $\mathcal H^{(1)}$. Then
\begin{equation}
\label{eq:3-bar-basis}
\big\{|f^k\rangle:=\varepsilon^{ijk}|e_i \wedge e_j \rangle\big\}_{1\leq k\leq 3},
\end{equation}
provides an orthonormal basis for $\mathcal H^{(2)}$. This choice of basis is particularly useful if we take into account
that it provides a basis for the ($SU(3)$) representation $\bar 3$ obtained from the decomposition
$3\otimes 3 = 6\oplus \bar 3$, corresponding to symmetrisation/antisymmetrisation of tensors in $\mathcal H^{(1)}\otimes \mathcal H^{(1)}$.  The
$|f^i\rangle$ span the $\bar 3$ representation.

Here, the representation $3$ stands for the defining   $U(3)$ representation on     $\mathcal H^{(1)}$  ($U^{(1)}(g)=g$).
This means that, at the two fermion level,
one particle observables are given  by the action of $\mathds C U(3)$  on $\mathcal H^{(2)}$. This action is obtained
from  the restriction of  operators of the form
$\widehat \alpha=   \int_{U(d)}d\mu(g)  \alpha(g) U^{(1)}(g)\otimes U^{(1)}(g)$
to $\bar 3$, regarded as a subspace of $3\times 3$.

It follows that
the algebra $\mathcal A$ of observables \emph{for the two-fermion system} is  generated by $|f^i\rangle\langle f^j|$ ($i,j=1,2,3$). Hence,  $\mathcal A \cong M_3(\mathds C)$.

Below we  consider two different choices  for the subalgebra $\mathcal A_0$. This will make clear that the notion of
\emph{entanglement} is very sensitive not only to the choice of state, but also to the choice of the observable algebra.

\noindent{\it {Choice 1 for $\mathcal A_0$ :}}

Here we consider $\mathcal A_0$ to be the full algebra $\mathcal A$. That is,  $\mathcal A_0$ is chosen to be the
full algebra of one-particle observables acting on $\mathcal H^{(2)}$. Now, we pick any two-fermion \emph{pure} state
$\omega: \mathcal A\rightarrow \mathds C$. Being a pure state on $\mathcal A \cong A_0$, the GNS representation corresponding to the pair $(\mathcal A_0, \omega)$ is irreducible. This is equivalent to the statement that
$\bar 3$ is an irreducible $SU(3)$ representation. This in turn  corresponds to the well-known fact that, for $\mathcal H^{(1)}= \mathds C^3$, all two-fermion vector states have Slater rank 1 (cf.~\cite{Ghirardi2004}). These states are therefore to be considered as non-entangled states. Notice, however, that if we use partial trace to compute the von Neumann entropy we get a result different from zero. In contrast, computing the von Neumann entropy via the GNS construction gives automatically zero in this case (because of irreducibility), this being in accordance with the fact that all states for $d=3$ are non-entangled (as long as $\mathcal A_0=\mathcal A$).

\noindent{\it {Choice 2 for $\mathcal A_0$ :}}

Now we let $\mathcal A_0$ be the subalgebra of  $\mathcal A$ consisting of all one-particle observables that involve
\emph{only} the one-particle states $|e_1\rangle$ and $|e_2\rangle$ and unity $\mathds 1_{\mathcal A}= \mathds 1_3$ . It can be easily checked that this subalgebra
is generated by the operators $M^{ij}:=|f^i\rangle\langle f^j|$ $(i,j = 1,2)$, as well as $\mathds 1_{\mathcal A}$. This is, therefore, a five dimensional matrix algebra.


Consider the following two-fermion vector state:
\begin{equation}
|\psi_\theta\rangle=   \cos\theta|f^1\rangle + \sin\theta |f^3\rangle,
\end{equation}
and let $\omega_\theta: \mathcal A  \rightarrow  \mathds C$ denote the corresponding state:
\begin{equation}
\omega_\theta (\alpha)= \langle\psi_\theta |\alpha|  \psi_\theta \rangle,\;\;\;\;\forall\alpha\in \mathcal A.
\end{equation}
Consider now the restriction of $\omega_\theta$ to the subalgebra $\mathcal A_0$:
\begin{equation}
\omega_{\theta,0}=\omega_\theta\mid_{\mathcal A_0}.
\end{equation}
We proceed to perform the GNS construction corresponding to the pair $\mathcal A_0,\omega_{\theta,0}$,
assuming that $0 < \theta < \frac{\pi}{2}$. By direct computation we check  that both $M^{12}$ and $M^{22}$ are null vectors:
\[
|[M^{12}]\rangle = |[M^{22}]\rangle= 0.
\]
In this range of values for $\theta$ these are all the linearly independent null vectors. This
can be seen from
\begin{equation}
 \langle\psi_\theta |\alpha^*\alpha | \psi_\theta \rangle = 0 \Rightarrow \alpha|\psi_\theta \rangle = 0
\Rightarrow \alpha = \sum c_iM^{i2}, c_i \in \mathds C.
\end{equation}
This means that the null space $ \mathcal N_{\theta,0}$ is two-dimensional and that, therefore,
$\mathcal H_\theta= \hat{\mathcal A_0}/\mathcal N_{\theta,0}$ is isomorphic to $\mathds C^3$,
 with basis
$\{|[M^{11}]\rangle, |[M^{21}]\rangle,|[E^3]\rangle\}$, where
$E^3:=\mathds 1_{\mathcal A}-M^{11}-M^{22}$.

Noticing that $\alpha_0 \in \mathcal A_0$ implies $\alpha_0\, E^3=0$,
we can check that the GNS space has the following decomposition in terms of irreducible representations:
$\mathcal H_\theta=\mathds C^2\oplus\mathds C^1$.

Denoting $P_1$ and $P_2$ the corresponding projections and using the fact that
 $[M^{11}+ M^{22}] ~=~ [\mathds 1_2]$, we obtain
\begin{equation}
P_1 |[  \mathds 1_{\mathcal A} ]\rangle=|[ M^{11} ]\rangle ,~~
P_2 |[  \mathds 1_{\mathcal A} ]\rangle=|[ E^3  ]\rangle
\end{equation}
Using the inner product of $\mathcal H_{\theta}$ to compute $|\mu_i|^2= \|P_i |[  \mathds 1_{\mathcal A} ]\rangle\|^2$, we obtain
\begin{equation}
|\mu_1|^2 = \cos^2\theta~,~~|\mu_2|^2 = \sin^2\theta.
\end{equation}
Hence,
\begin{equation}
\omega_{\theta,0}~=~\cos^2\theta\left(\frac{1}{\cos^2\theta} |[M^{11}]\rangle \langle[M^{11}]|\right)
+ \sin^2\theta\left(\frac{1}{\sin^2\theta} |[E^3]\rangle \langle[E^3]|\right).
\end{equation}
The result for the  entropy as a function of $\theta$ is therefore
\begin{equation}
\label{eq:S(theta)}
S(\theta) = -\cos^2 \theta \log \cos^2\theta - \sin^2 \theta \log \sin^2\theta.
\end{equation}

The cases $\theta=0$ and $\theta= \pi/2$ differ from the above mainly in the dimension of the GNS Hilbert space. Nevertheless,
the result for the entropy remains the same. It is given by the same formula (\ref{eq:S(theta)}), extended now to the values
$\theta=0$ and $\theta= \pi/2$. This means that the entropy for $\theta=0$ or $\theta=\pi/2$ is zero, and corresponds to the fact that, for these values of $\theta$, the GNS Hilbert space is irreducible.
Indeed, the result for the GNS space in the range $\theta\in [0,\pi/2 ]$ is:
\begin{eqnarray}
\mathcal H_\theta\cong \left\lbrace\begin{array}{cc}
\mathds C^2,&\;\theta=0\\
\mathds C^3\cong \mathds C^2\oplus \mathds C, &\; \theta\in (0,\pi/2)\\
\mathds C,\; &\theta=\pi/2.
\end{array}
\right.
\end{eqnarray}
This result should be contrasted against the fact that the $\bar 3$ representation,
when regarded as a representation space for  $SU(2)$,
splits as $2\oplus 1$.


 \subsubsection{{\bf Two Bosons, $\mathcal{H}^{(1)}=\mathds{C}^3$}}

This is the bosonic counterpart of the previous example.
Let $\{|e_1\rangle, |e_2\rangle,|e_3\rangle\}$ denote an orthonormal basis for $\mathcal H^{(1)}=\mathds C^3$.
 Recalling the decomposition $3\otimes 3 = 6\oplus \bar 3$ of the previous example, we see
 that the six-dimensional space corresponds to  the two-particle space $\mathcal H^{(2)}$, i.e.,
the space of  symmetric tensors in  $\mathcal H^{(1)}\otimes \mathcal H^{(1)}$.
The algebra $\mathcal A$ of observables for this two-boson system is then isomorphic to $M_6(\mathds C)$.
As a basis for $\mathcal H^{(2)}$ we choose the  vectors $\{|e_i \vee e_j\rangle \}_{i,j\in \{1,2,3\}}$, where
\begin{equation}
|e_i \vee e_j\rangle \equiv\left\{
\begin{array}{cc}
\frac{1}{\sqrt 2}(|e_i\rangle \otimes |e_j\rangle + |e_j\rangle \otimes |e_i\rangle), & i\neq j,\\
|e_i\rangle \otimes |e_i\rangle, &i =j.
\end{array}
\right.
\end{equation}
They form an orthonormal basis.

Now we consider the (pure) state $\omega_{(\theta,\phi)}: \mathcal A\rightarrow \mathds C$ that corresponds to
\begin{equation}
\label{eq:bosons}
|\psi_{(\theta,\phi)}\rangle = \sin\theta\cos\phi|e_1 \vee e_2\rangle +\sin\theta\sin\phi|e_1 \vee e_3\rangle + \cos\theta|e_3 \vee e_3\rangle.
\end{equation}

We are interested in the restriction of $\omega_{(\theta,\phi)}$ to the subalgebra $\mathcal A_0$ of one-particle observables
which besides $\mathds 1_6$, pertains \emph{only} to the one-particle vectors $|e_1\rangle$ and $|e_2\rangle$.
Proceeding in the same way as in the previous examples, we recognize that the 6 representation,
when regarded as a representation space for $SU(2)$ acting nontrivially on
$|e_1\rangle$ and $|e_2\rangle$, splits as $6=3\oplus 2\oplus 1$. The basis vectors for these three
invariant subspaces are given below:
\begin{eqnarray}
3:~~&|1\rangle = |e_1 \vee e_1\rangle,\; \; |0\rangle = |e_1 \vee e_2\rangle, \; \; |-1\rangle = |e_2 \vee e_2\rangle, &
\nonumber\\
2:~~&|1/2\rangle = |e_1 \vee e_3\rangle,\;\; |-1/2\rangle=|e_2 \vee e_3\rangle, \label{invariant} &\\
1:~~& |\tilde 0\rangle = |e_3 \vee e_3\rangle. &\nonumber
\end{eqnarray}
The one-particle observables on $\mathcal H^{(2)}$ are obtained
from  the operators $|e_i\rangle\langle e_j|$ (with $i,j=1,2$),
as well as from the unit operator on $\mathcal H^{(1)}$, by means of the coproduct.
Thus, the  subalgebra $\mathcal A_0$ is generated
by operators of the form $|u\rangle\langle v|$, with both $|u\rangle$ and $|v\rangle$ belonging to the \emph{same}
irreducible component of $\mathcal H^{(2)}$. (Note that the image of unity on $\mathcal H^{(1)}$ under the coproduct
$\Delta$ is $\mathds 1_{\mathcal A}$. Hence by taking combinations of images
of the above $\mathcal H^{(1)}$-observables under
$\Delta$, we see that $\mathcal A_0$ contains $|\tilde 0\rangle \langle \tilde 0|$).
In other words, $\mathcal A_0$ is given by block-diagonal matrices, with each block corresponding
to one of the irreducible components in the decomposition $6=3\oplus 2\oplus 1$. The dimension of
$\mathcal A_0$ is therefore $3^2+2^2+1^2 ~=~ 14$.

The construction of the GNS-representation corresponding to each particular value of the parameters
$\theta$ and $\phi$ is performed following the same procedure as in the previous examples. Let us introduce the
notation $B_{u, v}\equiv |u\rangle\langle v|$, for any pair $|u\rangle, |v\rangle $ in (\ref{invariant}). Then,
from (\ref{eq:bosons}) we see that as long as the $(\theta,\phi)$-coefficients are all different from zero,
those elements of $\mathcal A_0$ of the form $B_{j,\pm 1}$ ($j=0,\pm 1$) and $B_{\sigma,-1/2}$ ($\sigma=\pm 1/2$)
generate the null vectors. That these generate all the null vectors
follows from the fact that (\ref{eq:bosons}) contains
one basis element for every irreducible component,
so that no further linear relation can arise that lead to null vectors.
So in this case we have
\begin{equation}
\mathcal H_{(\theta,\phi)} : = {\hat{\mathcal A_0}}/{\mathcal N_{(\theta,\phi),0}} = \mathds C^6,
~~~~\mathcal N_{(\theta,\phi),0} = {\rm Null~space~}.
\end{equation}
In terms of irreducible subspaces, one can readily see that $\mathds C^6$ decomposes according to
$\mathds C^6 = \mathds C^3 \oplus \mathds C^2 \oplus \mathds C^1$.

In general, we can read off the decomposition of $\mathcal H_{(\theta,\phi)}$ into irreducible subspaces
from (\ref{eq:bosons}), depending on which of its coefficents vanish.
For example, if only the first one vanishes,
\begin{equation}
\mathcal H_{(\theta,\phi)} = \mathds C^2 \oplus \mathds C^1.
\end{equation}

It is interesting to consider  the entropy as a function of $(\theta,\phi)$. For the case
in which all $(\theta,\phi)$-coefficients are non-zero,  we have:
\begin{equation}
|[\mathds 1_{\mathcal A}]\rangle= |[B_{1,1}]\rangle+ |[B_{1/2,1/2}]\rangle + |[B_{\tilde 0, \tilde 0}]\rangle,
\end{equation}
from which the entropy is readily computed as before. The result is:
\begin{equation}
\label{eq:boson-entropy}
S(\theta,\phi) = -\sin^2\theta[\cos^2\phi\log (\sin \theta \cos\phi)^2 +
  \sin^2\phi \log (\sin \theta \sin\phi)^2] -
 \cos^2\theta\log (\cos\theta)^2.
\end{equation}

The analytic formulae for entropy when one or more of the coefficients in (\ref{invariant})
vanish can be obtained from
(\ref{eq:boson-entropy}) by taking suitable limits on $\theta$ and $\phi$.

\begin{figure}
\includegraphics[scale=.9]{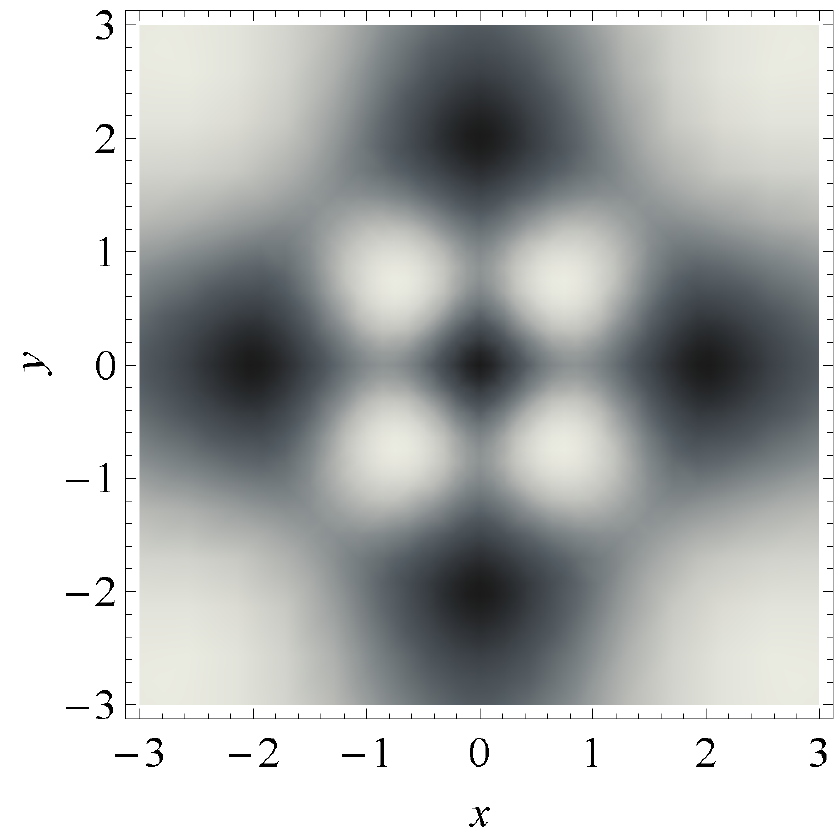}
\caption{\label{fig}
The entropy equation (\ref{eq:boson-entropy}) as a function of $x$ and $y$, the coordinates of a plane
representing the $(\theta,\phi)$-sphere through stereographic projection. Darker regions correspond to lower values of the entropy. Five of the six vanishing points of the entropy can be seen on the picture (black spots). The sixth one, corresponding to the north-pole of the sphere, lies `at infinity' in this representation.}
\end{figure}

We can see that the entropy vanishes whenever
$|\psi_{(\theta,\phi)}\rangle$ lies in a single irreducible component.
This happens precisely at those points of the two-sphere generated by the parameters $(\theta,\phi)$
that correspond to the coordinate axes. There are therefore six points where the entropy vanishes exactly.
This is depicted in Figure \ref{fig}, where the $(\theta,\phi)$-sphere has been mapped to the $x$-$y$ plane
through a stereographic projection. The figure shows the entropy as a function of the coordinates of
that plane.

\subsection{Entanglement for Braid Group Statistics}

Let us now outline how this approach can be used to compute entanglement entropy for systems with braid statistics. For that we need to recollect some facts from \cite{biedenharn1995quantum} of the quantum group $U_q(su(n))$.

\subsubsection{Preliminaries: Bosonic Realization of $U_q(su(n))$}

The $q$-number $[s]_q$ is defined by
\begin{equation}
  [s]_q=\frac{q^{\frac{s}{2}}-q^{-\frac{s}{2}}}{q^{\frac{1}{2}}-q^{-\frac{1}{2}}}.
\end{equation}
It satisfies the following properties of importance for us:
\begin{enumerate}
  \item $[s+t]_q=q^{-\frac{s}{2}} [t]_q+q^{\frac{t}{2}}[s]_q$.
  \item Jacobi identity: $[r]_q[s-t]_q+[s]_q[t-r]_q+[t]_q[r-s]_q=0$.
  \item $[1]_q=1$. By definition, $[0]_q=0$.
\end{enumerate}

Let $A,A^\dagger$ be $q$-deformed oscillators and let $a,a^\dagger$ be the standard undeformed bosonic oscillators. They are related by a dressing transformation:
\begin{align}
\label{dressing-1}
  A &= a\sqrt{\frac{[N]_q}{N}} = \sqrt{\frac{[N+1]_q}{N+1}} a, \\
  A^\dagger &= \sqrt{\frac{[N]_q}{N}} a^\dagger, \label{dressing-2}
\end{align}
where
\begin{equation}
N=a^\dagger a.
\end{equation}
From this dressing transformation we may find the representation for the $q$-deformed oscillators $A,A^\dagger$ from the known representation of $a,a^\dagger$. In particular, both sets have the same vacuum $|0\rangle$ such that $A|0\rangle=a|0\rangle=0$.

From (\ref{dressing-1}, \ref{dressing-2}), we see that
\begin{align}
 A^\dagger A &= [N]_q, A A^\dagger=[N+1]_q, \\
 N &= a^\dagger a
\end{align}
This gives
\begin{equation}
 AA^\dagger-q^{\frac{1}{2}} A^\dagger A =q^{-\frac{N}{2}}.
\end{equation}
Also
\begin{align}
  [N,A^\dagger]=A^\dagger, &\qquad [N,A]=-A; \nonumber \\
  [N,a^\dagger]=a^\dagger, &\qquad [N,a]=-a.
\end{align}

We may now construct $U_q(su(n))$ using a set of $N$ $A_i,A_i^\dagger$ oscillators with a fix $q$ following the Schwinger procedure. Let $\lambda_a$, with $a=1,..,n^2-1$, be the $n\times n$ Gell-Mann matrices of $su(n)$. Then
\begin{equation}
  \Lambda_a= A_i^\dagger (\lambda_a)^{ij} A_j
\end{equation}
are the generators of $U_q(su(n))$. We may organize this set of generators in the Cartan-Chevalley basis, where $H_i$, with $i=1,..,n-1$, generates the $U_q(su(n))$ Cartan subalgebra, and $E_{\pm\alpha}$ are ladder operators:
\begin{align}
  E_{ij}=A_i^\dagger A_j, &\qquad E_{ij}^\dagger = A_j^\dagger A_i, \quad \textrm{ for } i<j, \\
  H_l &= \frac{1}{2}\left( N_l-N_{l+1} \right), \quad \textrm{ for } l\leq n-1,
\end{align}
where
\begin{equation}
  N_l=a^\dagger_l a_l.
\end{equation}

In the case of $U_q(su(2))$, the Gell-Mann matrices are the $2\times 2$ Pauli matrices so that
\begin{equation}
  \Lambda_a= A_i^\dagger ~ (\sigma_a)^{ij} ~ A_j, \quad \textrm{ with }a=1,2,3 \textrm{ and } i=1,2.
\end{equation}
From this we obtain
\begin{align}
  E_{12}:=J_+=A_1^\dagger A_2, &\qquad E_{21}:=J_-=A_2^\dagger A_1, \nonumber \\
  H_1:= J_3 &= \frac{N_1-N_2}{2},
\end{align}
which satisfy
\begin{align}
  [J_3,J_\pm]=\pm J_\pm, \nonumber \\
  [J_+,J_-]=[2J_3]_q.
\end{align}

From now on, we consider only the cases for which $q$ is real and positive, $q>0$. In this case, $U_q(su(n))$ is a $*$-Hopf algebra with co-product
\begin{align}
\label{eq:q-coproduct}
  \Delta\left(J_\pm\right) &= q^{-\frac{J_3}{2}}\otimes J_\pm + J_\pm \otimes q^{\frac{J_3}{2}}, \nonumber \\
  \Delta\left(J_3\right) &= \mathds{1}\otimes J_3 + J_3 \otimes \mathds{1}.
\end{align}

The unitary irreducible representations (UIRR's) of $U_q(su(2))$ are labeled by $j\in\mathds{Z}^+/2$. For fixed $j$, $-j\leq m \leq j$, an orthonormal basis for the carrier vector space is
\begin{equation}
	\label{unirrep-q-basis}
  |jm\rangle = \frac{(A_1^\dagger)^{j+m} (A_2^\dagger)^{j-m}}{\sqrt{[j+m]_q![j-m]_q!}} |0\rangle,
\end{equation}
 where for $k\in\mathds{Z}^+$, $[k]_q!=[1]_q[2]_q...[k]_q$. The generators $J_\pm$ and $J_3$ act in (\ref{unirrep-q-basis}) as expected:
 \begin{align}
   J_\pm |jm\rangle &=  \sqrt{[j\mp m]_q[j\pm m + 1]_q}|j,m\pm 1\rangle, \\
   J_3 |jm\rangle &= m |jm\rangle.
 \end{align}

\subsubsection{The Braid Group}

We illustrate the ideas using the ``two-particle'' representation of $U_q(su(2))$.

For the undeformed oscillators, we can easily see that the vector state
$a_1^\dagger |0\rangle \otimes a_2^\dagger |0\rangle$ for example decomposes in terms of the action of the symmetric group $\mathcal{S}_2$ as
\begin{align}
a_1^\dagger |0\rangle \otimes a_2^\dagger |0\rangle &=\frac{1}{2}\left[ \left( a_1^\dagger |0\rangle \otimes a_2^\dagger |0\rangle + a_2^\dagger |0\rangle \otimes a_1^\dagger |0\rangle\right) + \left(a_1^\dagger |0\rangle \otimes a_2^\dagger |0\rangle - a_2^\dagger |0\rangle \otimes a_1^\dagger |0\rangle \right)\right], \nonumber \\ &\equiv |1,2\rangle_S+|1,2\rangle_A,
\end{align}
where indexes $S,A$ stand for symmetric and anti-symmetric, respectively.

For the $q$-deformed case, thi symmetric group decomposition has to be changed. This is simple to see.

We start with the highest weight state, call it $|1,1\rangle$, with respect to $U_q(su(2))$:
\begin{align}
  |1,1\rangle &=  A_1^\dagger |0\rangle \otimes A_1^\dagger |0\rangle \\
  \Delta(J_+) |1,1\rangle &= 0 \\
  \Delta(J_3) |1,1\rangle &= |1,1\rangle.
\end{align}
Next we lower using $J_-$:
\begin{align}
  \Delta(J_-) |1,1\rangle &= \left[ q^{-\frac{J_3}{2}}\otimes J_- + J_- \otimes q^{\frac{J_3}{2}}\right]~ A_1^\dagger |0\rangle \otimes A_1^\dagger |0\rangle \nonumber \\
  &= \left[q^{-\frac{1}{2}} A_1^\dagger |0\rangle \otimes A_2^\dagger |0\rangle + q^{\frac{1}{2}} A_2^\dagger |0\rangle \otimes A_1^\dagger |0\rangle \right].
\end{align}
It is thus clear that this state vector  cannot be decomposed under the symmetric group, since the $q$-powers are breaking the structure of the symmetric tensor product.

Formally, the structure we have just seen carries a representation of the braid group. We will not go into details here which can be found in \cite{biedenharn1995quantum, Balachandran2010}. In particular, as discussed by Biedenharn and Lohe, the multiparticle $q$-boson states constructed from the $q$-deformed oscillators are invariant under the braid group.

\subsubsection{Example}

The Schwinger realization of $q$-deformed oscillators given above is adapted to treat bosons. So we generalize the $q=1$ bosonic example above where $\mathcal{H}=\mathbb{C}^3$ and we observe just the algebra generated by the observables mixing  $1$ and $2$ and the unit operator.

Thus we now have three $q$-deformed oscillators $A_i,A_i^\dagger$ which commute for $i\neq j$. The two-particle $q$-boson states are spanned by $A_i^\dagger A_j^\dagger |0\rangle$ and is six-dimensional. Following (\ref{invariant}), we construct the orthonormal basis for the $3$ subspaces invariant under the observables acting on $1$ and $2$ particles and the unit operator:
\begin{align}
3:~~&|1\rangle_q = \frac{1}{\sqrt{[2]_q}}(A_1^\dagger)^2|0\rangle,\quad |0\rangle_q = A_1^\dagger A_2^\dagger|0\rangle, \quad |-1\rangle_q = \frac{1}{\sqrt{[2]_q}}(A_2^\dagger)^2|0\rangle,
\nonumber\\
2:~~&|1/2\rangle_q = A_1^\dagger A_3^\dagger |0\rangle, \qquad |-1/2\rangle_q=A_2^\dagger A_3^\dagger |0\rangle, \label{invariant-q} \\
1:~~& |\tilde 0\rangle_q = \frac{1}{\sqrt{[2]_q}}(A_3^\dagger)^2|0\rangle. \nonumber
\end{align}

We also generalize the vector state (\ref{eq:bosons}) to the normalized vector state
\begin{align}
  |\psi_{(\theta,\phi)}\rangle_q=\left(\sin\theta\cos\phi~A_1^\dagger A_2^\dagger+\sin\theta\sin\phi~A_1^\dagger A_3^\dagger +\cos\theta~\frac{1}{\sqrt{[2]_q}} (A_3^\dagger)^2 \right)|0\rangle.
\end{align}
It induces a state $\omega_{(\theta,\phi)}^{(q)}$ on the full algebra of $6\times 6$ matrices.

We now restrict $\omega_{(\theta,\phi)}^{(q)}$ to the observables pertaining to operators acting on $1$ and $2$ indices and the unit operator. This subalgebra $\mathcal{A}_0(q)$ is spanned by
\begin{equation}
	\label{subalgebra-1}
  |i\rangle_q ~_q\langle j|, \quad i,j\in [-1,0,1], \quad \textrm{and } \quad \mathds{1}_6.
\end{equation}

Now, the algebra $\mathcal{A}_0(q)$ generated by (\ref{subalgebra-1}) and the scalar products induced by $\omega_{(\theta,\phi)}^{(q)}$ are all independent of $q$. The conclusion is that the GNS Hilbert space and its properties are all quite independent of $q$. That includes entropy as well.

These observations can be generalized to more involved situations.

This simple example, to be contrasted with the usual two-fermion system already worked out here, shows once more that our approach allows one to naturally obtain a zero von Neumann entropy for separable systems even in the case of more sophisticated statistics. We think that this sets the stage for a more comprehensive study of systems with braid statistics, like Kitaev model, that may play a crucial role for developments of quantum computation.


\section{Time Evolution}
\label{sec:5}

If a unitary time evolution $U(t)$ of a pure state $\omega$ on an algebra $\mathcal A$ is given, then the time evolution of its restriction $\omega|_{\mathcal A_0} = \omega_0$ is determined by
\begin{equation}
\omega_0\rightarrow~\omega_0(t)=\left[U(t)~\omega\right]|_{\mathcal A_0},\;\;\;\omega_0(0)=\omega_0.
\end{equation}
The evolution of $\omega_0$ is in general by positive maps.
This fact is a consequence of the Stinespring-Choi theorem. That is the case even when $U(t)$ gives a unitary evolution on $\omega$ with Hamiltonian $H$:
\begin{equation}
U(t)~\omega=e^{iHt}~\omega~e^{-iHt}.
\end{equation}
Here $\omega$ is a density matrix.

An important point is that \emph{the rank of} $\omega_0(t)$ \emph{need not be continuous in} $t$ \emph{even if that of}
$U(t)~\omega$ \emph{is continuous in} $t$. \emph{It can change discontinuously}. This is shown by the example below. In that example, entropy is periodic in time, not monotone increasing in time, as is thought to be the case in nature.

The case of a fermion with $3$ internal degrees of freedom treated in section \ref{sec:identical-particles} is a simple example. The single particle Hilbert space $\mathcal H^{(1)}$ was $\mathds C^3$ with orthonormal basis
$\lbrace|e_i\rangle\rbrace_{i=1,2,3}$. The two-particle state space was $\bar 3= \bigwedge^2 \mathcal H^{(1)}\equiv\mathcal H^{(2)}$, with an orthonormal basis $\lbrace|f^i\rangle= \varepsilon^{ijk}|e_j\wedge e_k\rangle\rbrace_{i=1,2,3}$.

The single particle algebra acting on $\mathcal{H}^{(2)}$ was $\mathds{C}U(3)\otimes \mathds{1}_6$. We chose the pure state
\begin{equation}
\omega_\theta=|\psi_\theta\rangle\langle\psi_\theta|, \;\;\;|\psi_\theta\rangle=\cos\theta|f^1\rangle+\sin\theta|f^2\rangle,
\end{equation}
in the two-particle sector.

The subalgebra $\mathcal{A}_0$ was the image under the coproduct of the single particle algebra on $\mathcal H^{(1)}$  acting just on
$|e_1\rangle$ and
$|e_2\rangle$ .

Our results pertinent for the discussion of time evolution were
\begin{equation}
\omega_{\theta,0} = \cos^2 \theta~\rho^{(1)}_\theta + \sin^2 \theta~ \rho^{(3)}_\theta,
\end{equation}
where
\begin{align}
\rho^{(1)}_\theta &= \frac{1}{\cos^2\theta}|[M^{11}]\rangle\langle[M^{11}]|, \nonumber \\[0.2cm]
\rho^{(3)}_\theta &= \frac{1}{\sin^2\theta}|[E^{3}]\rangle\langle[M^{3}]|
\end{align}
and
\begin{equation}
\mathcal H_\theta^{\mbox{\tiny GNS}}=\left\lbrace \begin{array}{cc} \mathds C^2, & \theta=0,\\
\mathds C^3=\mathds C^2\oplus\mathds C, & 0<\theta<\pi/2,\\
\mathds C,& \theta=\pi/2.\end{array}\right.
\end{equation}
Thus the rank of $\omega_{\theta,0}=\omega_\theta|_{\mathcal A_0}$ jumps  from $2$ to $1$  as $\theta$ approaches $0$ or $\pi/2$.

Now consider the unitary evolution of $|\psi_\theta\rangle$ and $\omega_\theta$ under the self-adjoint  Hamiltonian
\begin{equation}
H=-i|f^2\rangle\langle f^1| + i |f^1\rangle\langle f^2|.
\end{equation}
It generates rotations in the plane of $\lbrace |f^1\rangle, |f^2\rangle\rbrace$ and hence changes $\theta$:
\begin{equation}
e^{itH}|\psi_\theta\rangle= \cos(\theta+ t)|f^1\rangle +  \sin(\theta+ t)|f^2\rangle.
\end{equation}
The restriction of this evolution  to $\omega_{\theta,0}$ is
\begin{equation}
U(t): \omega_{\theta,0}\rightarrow \omega_{\theta + t,0}.
\end{equation}
It is not unitary. It does not even preserve the rank of $\omega_{\theta,0}$: it jumps from $2$ to $1$ and back as $t$ increases.

We can write time evolution  as positive maps  so long as the rank of the density matrix  stays constant or decreases. Thus consider $\omega_{\theta,0}$ for $0<\theta<\pi/2$. It is of rank 2 expressible in terms of the orthonormal eigenvectors
\begin{equation}
|\chi^{(1)}(\theta)\rangle=\frac{1}{\cos\theta}|[M^{11}]\rangle, \qquad
|\chi^{(3)}(\theta)\rangle=\frac{1}{\sin\theta}|[M^{21}]\rangle
\end{equation}
and corresponding eigenvalues
\begin{equation}
\lambda_1(\theta) =\cos^2(\theta), \qquad \lambda_3(\theta) =\sin^2(\theta)
\end{equation}

For $0<\theta<\pi/2$, we can then write
\begin{equation}
\label{eq:theta-prime}
\omega_{\theta',0}=\sum_{a=1}^2 \Lambda_a^\dagger(\theta',\theta) \omega_{\theta,0} \Lambda_a(\theta',\theta),
\end{equation}
\begin{equation}
\Lambda_a(\theta',\theta)=\left(    \frac{\lambda_a(\theta')}{\lambda_a(\theta)}  \right)^{1/2}|\chi_a(\theta)\rangle\langle \chi_a(\theta')|.
\end{equation}
This makes sense for $\theta=0 (\pi/2)$ if the $a=2$ ($a=1$) term in  (\ref{eq:theta-prime}) is understood as zero.

But positive maps cannot increase the rank of a state. Hence we cannot write evolution starting from $\theta=0$ or $\pi/2$ in terms of positive maps.

\section{Anomalies and Restrictions}
\label{sec:6}

In this paper, mixed states emerge from restrictions of pure states $\omega$ on an algebra $\mathcal A$ to a subalgebra $\mathcal A_0$.

In a series of recent papers \cite{Balachandran2012,Balachandran2012b}, mixed states were introduced to eliminate anomalies. There it was proposed that anomalies can be eliminated by averaging say a pure state $\omega$ over the anomalous group.

We will now argue that the averaged  state in the second case can also be regarded as the restriction of $\omega$ to a subalgebra.

Let us focus on parity anomaly caused for example by QCD $\theta$-angle. The discussion is valid for any $\mathds{Z}_2$ symmetry group though.

Let $\mathcal A$ be the algebra of  observables with unity $\mathds 1$. If $P$ is parity, then $\mathcal A$ has the parity-even subalgebra
\begin{equation}
\mathcal A_+=\lbrace a_+\in \mathcal A: \;\;P a_+P=a_+\rbrace
\end{equation}
and also its complement
\begin{equation}
\mathcal A_-=\lbrace a_-\in \mathcal A: \;\;P a_-P=-a_-\rbrace
\end{equation}
which is not an algebra. The unity $\mathds 1$ is clearly in $\mathcal A_+$. But we assume that $\mathcal A_+$ has its own unity (projector) $\mathds 1_+$:
\begin{equation}
\mathds 1_+ a_+=a_+\mathds 1_+ = a_+,\;\;\mathds 1_+a _-=a_-\mathds 1_+=0.
\end{equation}

There is good physical meaning in assuming that $\mathcal A_+$ contains both $\mathds 1_+$ and $\mathds 1$. The projector $\mathds 1_+$ is the unity on the parity even elements. We need its orthogonal projector as observable
to tell us that the state of the system certainly has no component in $\mathcal A_+$. This orthogonal projector
is $\mathds 1_-$. And
$\mathds 1= \mathds 1_+ + \mathds 1_-$.

Then
\begin{equation}
\mathds 1_-= \mathds 1  -\mathds 1_+
\end{equation}
is the projector onto $\mathcal A_-$:
\begin{equation}
\mathds 1_- a_+=a_+\mathds 1_- =0 ,\;\;\mathds 1_-a _-=a_-\mathds 1_-=a_-.
\end{equation}
The parity-even subalgebra we consider is $\mathcal A_0 =\mathcal A_+\oplus \mathds C \mathds 1_-$

Let $\omega_\theta$, regarded as a density matrix, be such that
\begin{equation}
P\omega_\theta P= \omega_{-\theta}.
\end{equation}
For instance, a $\theta$-QCD state in QCD is of the above kind.

Since
\begin{equation}
\mathds 1= \mathds 1_+ + \mathds 1_-,
\end{equation}
$\omega_\theta$ splits  on restriction to $\mathcal A_0$
as
\begin{eqnarray}
\label{eq:6.9-theta}
\omega_\theta &=&|[\mathds 1_+]_{,\theta}\rangle\langle[\mathds 1_+]_{,\theta}|+|[\mathds 1_-]_{,\theta}\rangle\langle[\mathds 1_-]_{,\theta}|\nonumber\\
&=& \omega_{\theta,+} + \omega_{\theta,-}.
\end{eqnarray}
Now consider the expectation value $\omega_\theta(\alpha)$ for
\begin{equation}
\alpha= \alpha_++\lambda\mathds1_-\in \mathcal A_0.
\end{equation}
We have
\begin{equation}
\omega_\theta(\alpha):= \Tr~\omega_\theta \alpha = \omega_{\theta,+}(\alpha_+)+\omega_{\theta,-}(\lambda \mathds 1_-).
\end{equation}
Since
\begin{equation}
P\alpha_+P=\alpha_+,
\end{equation}
\begin{equation}
\omega_{\theta,+}(\alpha_+)=\omega_{-\theta,+}(\alpha_+)=\frac{1}{2}\left[\omega_{\theta,+}+\omega_{-\theta,+}\right](\alpha_+).
\end{equation}
Similarly since $P\mathds 1_- P=\mathds 1_-$,
\begin{equation}
\label{theta--1}
\omega_{\theta,-}=\frac{1}{2}\left[\omega_{\theta,-}+\omega_{-\theta,-}\right]
\end{equation}
Hence
\begin{equation}
\omega_\theta|_{\mathcal A_0}= \frac{1}{2}\left( \omega_\theta+\omega_{-\theta} \right)|_{\mathcal A_0}.
\end{equation}
But
\begin{equation}
(\omega_\theta+\omega_{-\theta}) (\alpha_-)=\omega_\theta(\alpha_-)+\omega_\theta(P\alpha_- P) = 0,
\end{equation}
which is true of RHS of (\ref{theta--1}) as well, extended to $\mathcal A_-$.

Hence \emph{restriction and averaging give same answer}.

\emph{They will coincide whenever $\mathcal A_0$  has a projector, the analogue of $\mathds 1_+$ here.} This is clear from the computations above. Note that the normalisation of vectors in (\ref{eq:6.9-theta}) is given by
\begin{equation}
\langle[\mathds 1_{\pm}]_{,\theta}|[\mathds 1_{\pm}]_{,\theta}\rangle= \omega_\theta(\mathds 1_\pm).
\end{equation}

\section{State Restrictions and quantum Observations}
\label{sec:7}

Suppose we study the restriction of a state $\omega$ on an algebra $\mathcal A$ of observables to a subalgebra $\mathcal A_0$. For reasons explained in section  \ref{sec:6}, we assume that $\mathcal A_0$ contains the projector $\mathds{1}_+$ and its orthogonal projector
\begin{equation}
\mathds{1}_- =\mathds 1 - \mathds{1}_+,
\end{equation}
$\mathds{1}$ being the unity of $\mathcal A$. The projectors satisfy
\begin{equation}
\mathds{1}_+\mathds{1}_- =0.
\end{equation}
Using these projectors, we can decompose $\mathcal A_0$ into two parts:
\begin{equation}
\mathcal A_0=\mathcal{A}_0\mathds{1}_+  \oplus \mathcal A_0 \mathds{1}_-.
\end{equation}
Let $\omega$ be pure on $\mathcal A$. Observe that $\mathcal{A}_0\mathds{1}_+$, $\mathcal{A}_0\mathds{1}_-$ are both invariant by $\mathcal A_0$. Then, in the GNS construction, the restricted state splits into two parts:
\begin{equation}
\label{eq:7.3observations}
\omega|_{\mathcal A_0}= |[\mathds{1}_+]\rangle \langle[\mathds{1}_+]|+|[\mathds{1}_-]\rangle \langle[\mathds{1}_-]|
\end{equation}
and is not pure.

Can we interpret (\ref{eq:7.3observations}) as emergent from observations?

The answer seems to be yes. If one measures the probability of finding either $1$ or $0$ for the observable $\mathds{1}_+$ on an ensemble with state $\omega$, the resultant state $\omega|_{\mathcal A_0}$ is after measurement exactly (\ref{eq:7.3observations}):
\begin{equation}
\omega ~\rightarrow~ \omega|_{\mathcal A_0}=\mathds{1}_+~|[\mathds 1]\rangle\langle[\mathds 1]|~\mathds{1}_+ + \mathds{1}_-~|[\mathds 1]\rangle\langle[\mathds 1]|~\mathds{1}_-  \label{eq:7-restriction}
\end{equation}
In this case, we are given the projector $\mathds{1}_+$ and we can reconstruct $\mathcal A_0\subseteq\mathcal A$ as its commutant:
\begin{equation}
\label{eq:7-commutant}
\mathcal A_0= \mbox{Commutant of $\mathds{1}_+$ in $\mathcal A$}.
\end{equation}
Working from this $\mathcal A_0$, we then show that the state restricted to (\ref{eq:7-commutant}) coincides
with (\ref{eq:7-restriction}).
\section{Conclusions}

We have seen in this work that there is a natural formulation of quantum physics dispensing with the use of Hilbert space as initial data which is well-adapted to the study of entanglement and entropy. In this approach, Hilbert space is an emergent concept. Instead, the initial data are the algebra of observables and their expectation values. From the expectation values one abstracts the notion of a state on the algebra.

In this formulation, the Hilbert space is obtained from the GNS construction that resembles the construction of the regular representation of finite  (or compact) groups. Furthermore, one may compute the von Neumann entropy associated with a density matrix that is obtained from a state on the algebra. It should be emphasized that for each state one may associate many distinct density matrices and therefore distinct von Neumann entropies. The discussion of this point is carried out in \cite{Balachandran2012c}.

A state $\omega$ on an algebra $\mathcal{A}$ can be restricted to a subalgebra $\mathcal{A}_0$. The new state $\omega|_{\mathcal{A}_0}$ may not be pure even if $\omega$ is. Its entropy is a measure of entanglement of $\mathcal{A}_0$ with $\mathcal{A}$.

This new approach to entanglement lets us treat identical particles obeying Bose, Fermi or even braid statistics with ease. Particle identity has posed severe problems in conventional approaches.

We have also shown how time evolution by positive maps for $\omega|_{\mathcal{A}_0}$ emerges when $\omega$ evolves unitarily.

Further points we have treated concerning quantum anomalies and their elimination by restricting states to subalgebras. In this manner, we can understand the use of mixed states to eliminate anomalies suggested by our previous work \cite{Balachandran2012}.  We also discussed how the restriction
$\omega|_{\mathcal{A}_0}$  emerges from a standard interpretation of  quantum physics from the observation of projectors.

\section*{Acknowledgments}

The authors would like to thank Alonso Botero for
discussions that led to this work. We also thank M. Asorey, B. Carneiro da Cunha, S. Ghosh, K. Gupta, A. Ibort, G. Marmo, V. P. Nair and A. Pinzul for
fruitful discussions during different stages of this work. APB is supported by the Institute of Mathematical Sciences, Chennai. ARQ is supported by CNPq under process number 307760/2009-0. AFRL is
supported by Universidad de los Andes.

\bibliography{GNS}

\begin{thebibliography}{28}%
\makeatletter
\providecommand \@ifxundefined [1]{%
 \@ifx{#1\undefined}
}%
\providecommand \@ifnum [1]{%
 \ifnum #1\expandafter \@firstoftwo
 \else \expandafter \@secondoftwo
 \fi
}%
\providecommand \@ifx [1]{%
 \ifx #1\expandafter \@firstoftwo
 \else \expandafter \@secondoftwo
 \fi
}%
\providecommand \natexlab [1]{#1}%
\providecommand \enquote  [1]{``#1''}%
\providecommand \bibnamefont  [1]{#1}%
\providecommand \bibfnamefont [1]{#1}%
\providecommand \citenamefont [1]{#1}%
\providecommand \href@noop [0]{\@secondoftwo}%
\providecommand \href [0]{\begingroup \@sanitize@url \@href}%
\providecommand \@href[1]{\@@startlink{#1}\@@href}%
\providecommand \@@href[1]{\endgroup#1\@@endlink}%
\providecommand \@sanitize@url [0]{\catcode `\\12\catcode `\$12\catcode
  `\&12\catcode `\#12\catcode `\^12\catcode `\_12\catcode `\%12\relax}%
\providecommand \@@startlink[1]{}%
\providecommand \@@endlink[0]{}%
\providecommand \url  [0]{\begingroup\@sanitize@url \@url }%
\providecommand \@url [1]{\endgroup\@href {#1}{\urlprefix }}%
\providecommand \urlprefix  [0]{URL }%
\providecommand \Eprint [0]{\href }%
\providecommand \doibase [0]{http://dx.doi.org/}%
\providecommand \selectlanguage [0]{\@gobble}%
\providecommand \bibinfo  [0]{\@secondoftwo}%
\providecommand \bibfield  [0]{\@secondoftwo}%
\providecommand \translation [1]{[#1]}%
\providecommand \BibitemOpen [0]{}%
\providecommand \bibitemStop [0]{}%
\providecommand \bibitemNoStop [0]{.\EOS\space}%
\providecommand \EOS [0]{\spacefactor3000\relax}%
\providecommand \BibitemShut  [1]{\csname bibitem#1\endcsname}%
\let\auto@bib@innerbib\@empty
\bibitem [{\citenamefont {Haag}(1996)}]{haag1996local}%
  \BibitemOpen
  \bibfield  {author} {\bibinfo {author} {\bibfnamefont {R.}~\bibnamefont
  {Haag}},\ }\href@noop {} {\emph {\bibinfo {title} {Local quantum physics}}}\
  (\bibinfo  {publisher} {Text and Monographs in Physics, 2nd ed., Springer},\
  \bibinfo {year} {1996})\BibitemShut {NoStop}%
\bibitem [{\citenamefont {Balachandran}\ \emph {et~al.}(2013)\citenamefont
  {Balachandran}, \citenamefont {Govindarajan}, \citenamefont {de~Queiroz},\
  and\ \citenamefont {Reyes-Lega}}]{balachandran2012a}%
  \BibitemOpen
  \bibfield  {author} {\bibinfo {author} {\bibfnamefont {A.~P.}\ \bibnamefont
  {Balachandran}}, \bibinfo {author} {\bibfnamefont {T.~R.}\ \bibnamefont
  {Govindarajan}}, \bibinfo {author} {\bibfnamefont {A.~R.}\ \bibnamefont
  {de~Queiroz}}, \ and\ \bibinfo {author} {\bibfnamefont {A.~F.}\ \bibnamefont
  {Reyes-Lega}},\ }\href@noop {} {\bibfield  {journal} {\bibinfo  {journal}
  {Phys. Rev. Lett. (in press)}\ } (\bibinfo {year} {2013})},\ \Eprint
  {http://arxiv.org/abs/1205.2882} {1205.2882} \BibitemShut {NoStop}%
\bibitem [{\citenamefont {Li}\ \emph {et~al.}(2001)\citenamefont {Li},
  \citenamefont {Zeng}, \citenamefont {Liu},\ and\ \citenamefont
  {Long}}]{Li2001}%
  \BibitemOpen
  \bibfield  {author} {\bibinfo {author} {\bibfnamefont {Y.~S.}\ \bibnamefont
  {Li}}, \bibinfo {author} {\bibfnamefont {B.}~\bibnamefont {Zeng}}, \bibinfo
  {author} {\bibfnamefont {X.~S.}\ \bibnamefont {Liu}}, \ and\ \bibinfo
  {author} {\bibfnamefont {G.~L.}\ \bibnamefont {Long}},\ }\href {\doibase
  10.1103/PhysRevA.64.054302} {\bibfield  {journal} {\bibinfo  {journal} {Phys.
  Rev. A}\ }\textbf {\bibinfo {volume} {64}},\ \bibinfo {pages} {054302}
  (\bibinfo {year} {2001})}\BibitemShut {NoStop}%
\bibitem [{\citenamefont {Paskauskas}\ and\ \citenamefont
  {You}(2001)}]{Paskauskas2001}%
  \BibitemOpen
  \bibfield  {author} {\bibinfo {author} {\bibfnamefont {R.}~\bibnamefont
  {Paskauskas}}\ and\ \bibinfo {author} {\bibfnamefont {L.}~\bibnamefont
  {You}},\ }\href {\doibase 10.1103/PhysRevA.64.042310} {\bibfield  {journal}
  {\bibinfo  {journal} {Phys. Rev. A}\ }\textbf {\bibinfo {volume} {64}},\
  \bibinfo {pages} {042310} (\bibinfo {year} {2001})}\BibitemShut {NoStop}%
\bibitem [{\citenamefont {Schliemann}\ \emph {et~al.}(2001)\citenamefont
  {Schliemann}, \citenamefont {Cirac}, \citenamefont {Ku\ifmmode~\acute{s}\else
  \'{s}\fi{}}, \citenamefont {Lewenstein},\ and\ \citenamefont
  {Loss}}]{Schliemann2001}%
  \BibitemOpen
  \bibfield  {author} {\bibinfo {author} {\bibfnamefont {J.}~\bibnamefont
  {Schliemann}}, \bibinfo {author} {\bibfnamefont {J.~I.}\ \bibnamefont
  {Cirac}}, \bibinfo {author} {\bibfnamefont {M.}~\bibnamefont
  {Ku\ifmmode~\acute{s}\else \'{s}\fi{}}}, \bibinfo {author} {\bibfnamefont
  {M.}~\bibnamefont {Lewenstein}}, \ and\ \bibinfo {author} {\bibfnamefont
  {D.}~\bibnamefont {Loss}},\ }\href {\doibase 10.1103/PhysRevA.64.022303}
  {\bibfield  {journal} {\bibinfo  {journal} {Phys. Rev. A}\ }\textbf {\bibinfo
  {volume} {64}},\ \bibinfo {pages} {022303} (\bibinfo {year}
  {2001})}\BibitemShut {NoStop}%
\bibitem [{\citenamefont {Zanardi}(2002)}]{Zanardi2002}%
  \BibitemOpen
  \bibfield  {author} {\bibinfo {author} {\bibfnamefont {P.}~\bibnamefont
  {Zanardi}},\ }\href {\doibase 10.1103/PhysRevA.65.042101} {\bibfield
  {journal} {\bibinfo  {journal} {Phys. Rev. A}\ }\textbf {\bibinfo {volume}
  {65}},\ \bibinfo {pages} {042101} (\bibinfo {year} {2002})}\BibitemShut
  {NoStop}%
\bibitem [{\citenamefont {Fang}\ and\ \citenamefont {Chang}(2003)}]{Fang2003}%
  \BibitemOpen
  \bibfield  {author} {\bibinfo {author} {\bibfnamefont {A.}~\bibnamefont
  {Fang}}\ and\ \bibinfo {author} {\bibfnamefont {Y.}~\bibnamefont {Chang}},\
  }\href {\doibase 10.1016/S0375-9601(03)00546-2} {\bibfield  {journal}
  {\bibinfo  {journal} {Physics Letters A}\ }\textbf {\bibinfo {volume}
  {311}},\ \bibinfo {pages} {443–} (\bibinfo {year} {2003})}\BibitemShut
  {NoStop}%
\bibitem [{\citenamefont {Ghirardi}\ and\ \citenamefont
  {Marinatto}(2004)}]{Ghirardi2004}%
  \BibitemOpen
  \bibfield  {author} {\bibinfo {author} {\bibfnamefont {G.}~\bibnamefont
  {Ghirardi}}\ and\ \bibinfo {author} {\bibfnamefont {L.}~\bibnamefont
  {Marinatto}},\ }\href {\doibase 10.1103/PhysRevA.70.012109} {\bibfield
  {journal} {\bibinfo  {journal} {Phys. Rev. A}\ }\textbf {\bibinfo {volume}
  {70}},\ \bibinfo {pages} {012109} (\bibinfo {year} {2004})}\BibitemShut
  {NoStop}%
\bibitem [{\citenamefont {L\'evay}\ \emph {et~al.}(2005)\citenamefont
  {L\'evay}, \citenamefont {Nagy},\ and\ \citenamefont {Pipek}}]{L'evay2005}%
  \BibitemOpen
  \bibfield  {author} {\bibinfo {author} {\bibfnamefont {P.}~\bibnamefont
  {L\'evay}}, \bibinfo {author} {\bibfnamefont {S.}~\bibnamefont {Nagy}}, \
  and\ \bibinfo {author} {\bibfnamefont {J.}~\bibnamefont {Pipek}},\ }\href
  {\doibase 10.1103/PhysRevA.72.022302} {\bibfield  {journal} {\bibinfo
  {journal} {Phys. Rev. A}\ }\textbf {\bibinfo {volume} {72}},\ \bibinfo
  {pages} {022302} (\bibinfo {year} {2005})}\BibitemShut {NoStop}%
\bibitem [{\citenamefont {Bañuls}\ \emph {et~al.}(2009)\citenamefont {Bañuls},
  \citenamefont {Cirac},\ and\ \citenamefont {Wolf}}]{Banuls2009}%
  \BibitemOpen
  \bibfield  {author} {\bibinfo {author} {\bibfnamefont {M.-C.}\ \bibnamefont
  {Bañuls}}, \bibinfo {author} {\bibfnamefont {J.~I.}\ \bibnamefont {Cirac}}, \
  and\ \bibinfo {author} {\bibfnamefont {M.~M.}\ \bibnamefont {Wolf}},\ }\href
  {\doibase 10.1088/1742-6596/171/1/012032} {\bibfield  {journal} {\bibinfo
  {journal} {Journal of Physics: Conference Series}\ }\textbf {\bibinfo
  {volume} {171}},\ \bibinfo {pages} {012032} (\bibinfo {year}
  {2009})}\BibitemShut {NoStop}%
\bibitem [{\citenamefont {Plastino}\ \emph {et~al.}(2009)\citenamefont
  {Plastino}, \citenamefont {Manzano},\ and\ \citenamefont
  {Dehesa}}]{Plastino2009}%
  \BibitemOpen
  \bibfield  {author} {\bibinfo {author} {\bibfnamefont {A.}~\bibnamefont
  {Plastino}}, \bibinfo {author} {\bibfnamefont {D.}~\bibnamefont {Manzano}}, \
  and\ \bibinfo {author} {\bibfnamefont {J.}~\bibnamefont {Dehesa}},\ }\href
  {\doibase 10.1209/0295-5075/86/20005} {\bibfield  {journal} {\bibinfo
  {journal} {European Physics Letters}\ }\textbf {\bibinfo {volume} {86}},\
  \bibinfo {pages} {20005} (\bibinfo {year} {2009})}\BibitemShut {NoStop}%
\bibitem [{\citenamefont {Zander}\ \emph {et~al.}(2012)\citenamefont {Zander},
  \citenamefont {Plastino}, \citenamefont {Casas},\ and\ \citenamefont
  {Plastino}}]{Zander2012}%
  \BibitemOpen
  \bibfield  {author} {\bibinfo {author} {\bibfnamefont {C.}~\bibnamefont
  {Zander}}, \bibinfo {author} {\bibfnamefont {A.}~\bibnamefont {Plastino}},
  \bibinfo {author} {\bibfnamefont {M.}~\bibnamefont {Casas}}, \ and\ \bibinfo
  {author} {\bibfnamefont {A.}~\bibnamefont {Plastino}},\ }\href
  {http://dx.doi.org/10.1140/epjd/e2011-10654-x} {\bibfield  {journal}
  {\bibinfo  {journal} {The European Physical Journal D - Atomic, Molecular,
  Optical and Plasma Physics}\ }\textbf {\bibinfo {volume} {66}},\ \bibinfo
  {pages} {1} (\bibinfo {year} {2012})},\ \bibinfo {note}
  {10.1140/epjd/e2011-10654-x}\BibitemShut {NoStop}%
\bibitem [{\citenamefont {Barnum}\ \emph {et~al.}(2004)\citenamefont {Barnum},
  \citenamefont {Knill}, \citenamefont {Ortiz}, \citenamefont {Rolando},\ and\
  \citenamefont {Viola}}]{Barnum2004}%
  \BibitemOpen
  \bibfield  {author} {\bibinfo {author} {\bibfnamefont {H.}~\bibnamefont
  {Barnum}}, \bibinfo {author} {\bibfnamefont {E.}~\bibnamefont {Knill}},
  \bibinfo {author} {\bibfnamefont {G.}~\bibnamefont {Ortiz}}, \bibinfo
  {author} {\bibfnamefont {S.}~\bibnamefont {Rolando}}, \ and\ \bibinfo
  {author} {\bibfnamefont {L.}~\bibnamefont {Viola}},\ }\href {\doibase
  10.1103/PhysRevLett.92.107902} {\bibfield  {journal} {\bibinfo  {journal}
  {Physical Review Letters}\ }\textbf {\bibinfo {volume} {92}},\ \bibinfo
  {pages} {107902} (\bibinfo {year} {2004})}\BibitemShut {NoStop}%
\bibitem [{\citenamefont {Benatti}\ \emph {et~al.}(2012)\citenamefont
  {Benatti}, \citenamefont {Floreanini},\ and\ \citenamefont
  {Marzolino}}]{Benatti2012}%
  \BibitemOpen
  \bibfield  {author} {\bibinfo {author} {\bibfnamefont {F.}~\bibnamefont
  {Benatti}}, \bibinfo {author} {\bibfnamefont {R.}~\bibnamefont {Floreanini}},
  \ and\ \bibinfo {author} {\bibfnamefont {U.}~\bibnamefont {Marzolino}},\
  }\href {\doibase 10.1016/j.aop.2012.02.002} {\bibfield  {journal} {\bibinfo
  {journal} {Annals of Physics}\ }\textbf {\bibinfo {volume} {327}},\ \bibinfo
  {pages} {1304 } (\bibinfo {year} {2012})}\BibitemShut {NoStop}%
\bibitem [{\citenamefont {Harshman}\ and\ \citenamefont
  {Ranade}(2011)}]{Harshman2011}%
  \BibitemOpen
  \bibfield  {author} {\bibinfo {author} {\bibfnamefont {N.~L.}\ \bibnamefont
  {Harshman}}\ and\ \bibinfo {author} {\bibfnamefont {K.~S.}\ \bibnamefont
  {Ranade}},\ }\href {\doibase 10.1103/PhysRevA.84.012303} {\bibfield
  {journal} {\bibinfo  {journal} {Phys. Rev. A}\ }\textbf {\bibinfo {volume}
  {84}},\ \bibinfo {pages} {012303} (\bibinfo {year} {2011})}\BibitemShut
  {NoStop}%
\bibitem [{\citenamefont {Derkacz}\ \emph {et~al.}(2012)\citenamefont
  {Derkacz}, \citenamefont {Gw{\'o}{\'z}d{\'z}},\ and\ \citenamefont
  {Jak{\'o}bczyk}}]{Derkacz2012}%
  \BibitemOpen
  \bibfield  {author} {\bibinfo {author} {\bibfnamefont {{\L}.}~\bibnamefont
  {Derkacz}}, \bibinfo {author} {\bibfnamefont {M.}~\bibnamefont
  {Gw{\'o}{\'z}d{\'z}}}, \ and\ \bibinfo {author} {\bibfnamefont
  {L.}~\bibnamefont {Jak{\'o}bczyk}},\ }\href@noop {} {\bibfield  {journal}
  {\bibinfo  {journal} {Journal of Physics A: Mathematical and Theoretical}\
  }\textbf {\bibinfo {volume} {45}},\ \bibinfo {pages} {025302} (\bibinfo
  {year} {2012})}\BibitemShut {NoStop}%
\bibitem [{\citenamefont {Balachandran}\ \emph {et~al.}(2010)\citenamefont
  {Balachandran}, \citenamefont {Jo},\ and\ \citenamefont
  {Marmo}}]{Balachandran2010}%
  \BibitemOpen
  \bibfield  {author} {\bibinfo {author} {\bibfnamefont {A.~P.}\ \bibnamefont
  {Balachandran}}, \bibinfo {author} {\bibfnamefont {S.~G.}\ \bibnamefont
  {Jo}}, \ and\ \bibinfo {author} {\bibfnamefont {G.}~\bibnamefont {Marmo}},\
  }\href@noop {} {\emph {\bibinfo {title} {Group Theory and Hopf Algebras:
  Lectures for Physicists}}}\ (\bibinfo  {publisher} {World Scientific},\
  \bibinfo {year} {2010})\BibitemShut {NoStop}%
\bibitem [{\citenamefont {Stinespring}(1955)}]{Stinespring1955}%
  \BibitemOpen
  \bibfield  {author} {\bibinfo {author} {\bibfnamefont {W.~F.}\ \bibnamefont
  {Stinespring}},\ }\href@noop {} {\bibfield  {journal} {\bibinfo  {journal}
  {Proc. Amer. Math. Soc.}\ }\textbf {\bibinfo {volume} {6}},\ \bibinfo {pages}
  {211} (\bibinfo {year} {1955})}\BibitemShut {NoStop}%
\bibitem [{\citenamefont {Balachandran}\ and\ \citenamefont
  {de~Queiroz}(2012)}]{Balachandran2012}%
  \BibitemOpen
  \bibfield  {author} {\bibinfo {author} {\bibfnamefont {A.~P.}\ \bibnamefont
  {Balachandran}}\ and\ \bibinfo {author} {\bibfnamefont {A.~R.}\ \bibnamefont
  {de~Queiroz}},\ }\href {\doibase 10.1103/PhysRevD.85.025017} {\bibfield
  {journal} {\bibinfo  {journal} {Phys. Rev. D}\ }\textbf {\bibinfo {volume}
  {85}},\ \bibinfo {pages} {025017} (\bibinfo {year} {2012})}\BibitemShut
  {NoStop}%
\bibitem [{\citenamefont {Balachandran}\ \emph
  {et~al.}(2012{\natexlab{a}})\citenamefont {Balachandran}, \citenamefont
  {Govindarajan},\ and\ \citenamefont {Queiroz}}]{Balachandran2012b}%
  \BibitemOpen
  \bibfield  {author} {\bibinfo {author} {\bibfnamefont {A.~P.}\ \bibnamefont
  {Balachandran}}, \bibinfo {author} {\bibfnamefont {T.~R.}\ \bibnamefont
  {Govindarajan}}, \ and\ \bibinfo {author} {\bibfnamefont {A.}~\bibnamefont
  {Queiroz}},\ }\href {\doibase 10.1140/epjp/i2012-12118-7} {\bibfield
  {journal} {\bibinfo  {journal} {The European Physical Journal Plus}\ }\textbf
  {\bibinfo {volume} {127}},\ \bibinfo {pages} {1} (\bibinfo {year}
  {2012}{\natexlab{a}})}\BibitemShut {NoStop}%
\bibitem [{\citenamefont {Sorkin}(2012)}]{Sorkin2012}%
  \BibitemOpen
  \bibfield  {author} {\bibinfo {author} {\bibfnamefont {R.~D.}\ \bibnamefont
  {Sorkin}},\ }\href@noop {} {\  (\bibinfo {year} {2012})},\ \Eprint
  {http://arxiv.org/abs/1205.2953} {arXiv:1205.2953 [hep-th]} \BibitemShut
  {NoStop}%
\bibitem [{\citenamefont {Balachandran}\ \emph
  {et~al.}(2012{\natexlab{b}})\citenamefont {Balachandran}, \citenamefont
  {de~Queiroz},\ and\ \citenamefont {Vaidya}}]{Balachandran2012c}%
  \BibitemOpen
  \bibfield  {author} {\bibinfo {author} {\bibfnamefont {A.~P.}\ \bibnamefont
  {Balachandran}}, \bibinfo {author} {\bibfnamefont {A.~R.}\ \bibnamefont
  {de~Queiroz}}, \ and\ \bibinfo {author} {\bibfnamefont {S.}~\bibnamefont
  {Vaidya}},\ }\href@noop {} {\  (\bibinfo {year} {2012}{\natexlab{b}})},\
  \Eprint {http://arxiv.org/abs/1212.1239} {arXiv:1212.1239 [hep-th]}
  \BibitemShut {NoStop}%
\bibitem [{\citenamefont {Landi}(2008)}]{Landi2008}%
  \BibitemOpen
  \bibfield  {author} {\bibinfo {author} {\bibfnamefont {G.}~\bibnamefont
  {Landi}},\ }\href@noop {} {\emph {\bibinfo {title} {An Introduction to
  Noncommutative Spaces and their Geometry}}}\ (\bibinfo  {publisher} {vol. 51,
  Lecture Notes in Physics Monographs, Springer-Verlag},\ \bibinfo {year}
  {2008})\BibitemShut {NoStop}%
\bibitem [{\citenamefont {Aneziris}\ \emph {et~al.}(1989)\citenamefont
  {Aneziris}, \citenamefont {Balachandran}, \citenamefont {Bourdeau},
  \citenamefont {Jo}, \citenamefont {Ramadas},\ and\ \citenamefont
  {Sorkin}}]{Aneziris1989}%
  \BibitemOpen
  \bibfield  {author} {\bibinfo {author} {\bibfnamefont {C.}~\bibnamefont
  {Aneziris}}, \bibinfo {author} {\bibfnamefont {A.~P.}\ \bibnamefont
  {Balachandran}}, \bibinfo {author} {\bibfnamefont {M.}~\bibnamefont
  {Bourdeau}}, \bibinfo {author} {\bibfnamefont {S.}~\bibnamefont {Jo}},
  \bibinfo {author} {\bibfnamefont {T.~R.}\ \bibnamefont {Ramadas}}, \ and\
  \bibinfo {author} {\bibfnamefont {R.~D.}\ \bibnamefont {Sorkin}},\ }\href
  {\doibase 10.1142/S0217751X8900234X} {\bibfield  {journal} {\bibinfo
  {journal} {Int.J.Mod.Phys.}\ }\textbf {\bibinfo {volume} {A4}},\ \bibinfo
  {pages} {5459} (\bibinfo {year} {1989})}\BibitemShut {NoStop}%
\bibitem [{\citenamefont {Aneziris}\ \emph {et~al.}(1991)\citenamefont
  {Aneziris}, \citenamefont {Balachandran}, \citenamefont {Kauffman},\ and\
  \citenamefont {Srivastava}}]{Aneziris1991}%
  \BibitemOpen
  \bibfield  {author} {\bibinfo {author} {\bibfnamefont {C.}~\bibnamefont
  {Aneziris}}, \bibinfo {author} {\bibfnamefont {A.~P.}\ \bibnamefont
  {Balachandran}}, \bibinfo {author} {\bibfnamefont {L.}~\bibnamefont
  {Kauffman}}, \ and\ \bibinfo {author} {\bibfnamefont {A.~M.}\ \bibnamefont
  {Srivastava}},\ }\href {\doibase 10.1142/S0217751X91001210} {\bibfield
  {journal} {\bibinfo  {journal} {Int.J.Mod.Phys.}\ }\textbf {\bibinfo {volume}
  {A6}},\ \bibinfo {pages} {2519} (\bibinfo {year} {1991})}\BibitemShut
  {NoStop}%
\bibitem [{\citenamefont {Balachandran}\ and\ \citenamefont
  {de~Queiroz}(2011)}]{Balachandran2011}%
  \BibitemOpen
  \bibfield  {author} {\bibinfo {author} {\bibfnamefont {A.~P.}\ \bibnamefont
  {Balachandran}}\ and\ \bibinfo {author} {\bibfnamefont {A.~R.}\ \bibnamefont
  {de~Queiroz}},\ }\href {\doibase 10.1007/JHEP11(2011)126} {\bibfield
  {journal} {\bibinfo  {journal} {JHEP}\ }\textbf {\bibinfo {volume} {1111}},\
  \bibinfo {pages} {126} (\bibinfo {year} {2011})},\ \Eprint
  {http://arxiv.org/abs/1109.5290} {arXiv:1109.5290 [hep-th]} \BibitemShut
  {NoStop}%
\bibitem [{\citenamefont {Tichy}\ \emph {et~al.}(2011)\citenamefont {Tichy},
  \citenamefont {Mintert},\ and\ \citenamefont {Buchleitner}}]{Tichy2011}%
  \BibitemOpen
  \bibfield  {author} {\bibinfo {author} {\bibfnamefont {M.~C.}\ \bibnamefont
  {Tichy}}, \bibinfo {author} {\bibfnamefont {F.}~\bibnamefont {Mintert}}, \
  and\ \bibinfo {author} {\bibfnamefont {A.}~\bibnamefont {Buchleitner}},\
  }\href {http://stacks.iop.org/0953-4075/44/i=19/a=192001} {\bibfield
  {journal} {\bibinfo  {journal} {Journal of Physics B: Atomic, Molecular and
  Optical Physics}\ }\textbf {\bibinfo {volume} {44}},\ \bibinfo {pages}
  {192001} (\bibinfo {year} {2011})}\BibitemShut {NoStop}%
\bibitem [{\citenamefont {Biedenharn}\ and\ \citenamefont
  {Lohe}(1995)}]{biedenharn1995quantum}%
  \BibitemOpen
  \bibfield  {author} {\bibinfo {author} {\bibfnamefont {L.~C.}\ \bibnamefont
  {Biedenharn}}\ and\ \bibinfo {author} {\bibfnamefont {M.~A.}\ \bibnamefont
  {Lohe}},\ }\href@noop {} {\emph {\bibinfo {title} {{Quantum Group Symmetry
  and q-Tensor Algebras}}}}\ (\bibinfo  {publisher} {World Scientific},\
  \bibinfo {year} {1995})\BibitemShut {NoStop}%
\end{thebibliography}%

\end{document}